\documentclass[12pt]{article}
\usepackage[dvips]{graphicx}
\usepackage{epsfig,cite}
\usepackage {amsmath}
\usepackage{amssymb}
\usepackage{slashed}
\newcount\timecount
\newcount\hours \newcount\minutes  \newcount\temp \newcount\pmhours
\hours = \time
\divide\hours by 60
\temp = \hours
\multiply\temp by 60
\minutes = \time
\advance\minutes by -\temp
\def\hour{\the\hours}
\def\minute{\ifnum\minutes<10 0\the\minutes
            \else\the\minutes\fi}
\def\clock{
\ifnum\hours=0 12:\minute\ AM
\else\ifnum\hours<12 \hour:\minute\ AM
      \else\ifnum\hours=12 12:\minute\ PM
            \else\ifnum\hours>12
                 \pmhours=\hours
                 \advance\pmhours by -12
                 \the\pmhours:\minute\ PM
                 \fi
            \fi
      \fi
\fi
}

\def\monthname{\relax\ifcase\month 0/\or January\or February\or
   March\or April\or May\or June\or July\or August\or September\or
   October\or November\or December\else\number\month/\fi}

\def\bold#1{\setbox0=\hbox{$#1$}%
     \kern-.025em\copy0\kern-\wd0
     \kern.05em\copy0\kern-\wd0
     \kern-.025em\raise.0433em\box0 }

\def\beq{\begin{equation}}
\def\eeq{\end{equation}}

\def\ga{\mathrel{\raise.3ex\hbox{$>$\kern-.75em\lower1ex\hbox{$\sim$}}}}
\def\la{\mathrel{\raise.3ex\hbox{$<$\kern-.75em\lower1ex\hbox{$\sim$}}}}
\def\gev{{\rm \, Ge\kern-0.125em V}}
\def\tev{{\rm \, Te\kern-0.125em V}}
\def\gyr{{\rm \, G\kern-0.125em yr}}

\def\gappeq{\mathrel{\rlap {\raise.5ex\hbox{$>$}}
{\lower.5ex\hbox{$\sim$}}}}
\def\lappeq{\mathrel{\rlap{\raise.5ex\hbox{$<$}}
{\lower.5ex\hbox{$\sim$}}}}
\def\Toprel#1\over#2{\mathrel{\mathop{#2}\limits^{#1}}}

\def\m12{m_{1\!/2}}

\newcommand\iso[2]{\mbox{${}^{#2}${\rm #1}}}
\def\he#1{\iso{He}{#1}}
\def\be#1{\iso{Be}{#1}}
\def\li#1{\iso{Li}{#1}}

\def\bea{\begin{eqnarray}}
\def\eea{\end{eqnarray}}

\def\beqar{\begin{eqnarray}}
\def\eeqar{\end{eqnarray}}

\usepackage{lscape}
\usepackage[dvips]{graphicx}

\def\beq{\begin{equation}}
\def\eeq{\end{equation}}

\begin{document}

\begin{titlepage}
\pagestyle{empty}
\rightline{KCL-PH/2013-08, LCTS/2013-04, CERN-PH-TH/2013-024}
\rightline{UMN--TH--3138/13, FTPI--MINN--13/05}
\vskip +0.4in
\begin{center}
{\large {\bf Gravitino Decays and the Cosmological Lithium Problem \\
\vskip 0.1in
in Light of the LHC Higgs and Supersymmetry Searches}}

\end{center}
\begin{center}
\vskip +0.25in
{\bf Richard~H.~Cyburt}$^{1}$, {\bf John~Ellis}$^{2,3}$,
{\bf Brian~D.~Fields}$^{4}$, \\
\vskip 0.1in
{\bf Feng Luo}$^{2}$, {\bf Keith~A.~Olive}$^{5}$
and {\bf Vassilis~C.~Spanos}$^{6}$\\
\vskip 0.2in
{\small {\it
$^1${Joint Institute for Nuclear Astrophysics (JINA), National Superconducting
Cyclotron Laboratory (NSCL), Michigan State University, East Lansing,
MI 48824, USA}\\ 
$^2${Theoretical Physics and Cosmology Group, Department of Physics, King's College London, London~WC2R 2LS, UK}\\
$^3${TH Division, Physics Department, CERN, CH-1211 Geneva 23, Switzerland}\\
$^4${Departments of Astronomy and of Physics, \\ University of Illinois, Urbana, IL 61801, USA}\\
$^5${William I. Fine Theoretical Physics Institute, School of Physics and Astronomy,\\
University of Minnesota, Minneapolis, MN 55455,\,USA}\\
$^6${Institute of Nuclear and Particle  Physics, NCSR ``Demokritos", GR-15310 Athens, Greece}} \\
}
\vspace{0.8cm}
{\bf Abstract}
\end{center}
{\small
We studied previously the impact on light-element
abundances of gravitinos decaying during or after Big-Bang nucleosynthesis (BBN). We found
regions of the gravitino mass $m_{3/2}$ and abundance $\zeta_{3/2}$ plane where its decays could
reconcile the calculated abundance of \li7 with observation without perturbing
the other light-element abundances unacceptably. Here we revisit this issue
in light of LHC measurements of the Higgs mass and constraints on supersymmetric model parameters,
as well as updates in the astrophysical measurements of light-element abundances. 
In addition to the constrained minimal supersymmetric extension of the Standard Model with universal soft supersymmetry-breaking masses
at the GUT scale (the CMSSM) studied previously, we also study models with universality imposed below the GUT
scale and models with non-universal Higgs masses (NUHM1). We calculate the total likelihood function 
for the light-element abundances, taking into account the observational uncertainties.
We find that gravitino decays provide a robust solution to the cosmological \li7
problem along strips in the $(m_{3/2}, \zeta_{3/2})$ plane along which the
abundances of deuterium, \he4 and \li7 may be fit with $\chi^2_{\rm min} \lappeq 3$,
compared with $\chi^2 \sim 34$ if the effects of gravitino decays are unimportant.
The minimum of the likelihood function is reduced to $\chi^2 < 2$ when the uncertainty on D/H is relaxed
and $< 1$ when the lithium abundance is taken from globular cluster data.}


\vfill
\leftline{March  2013}
\end{titlepage}

\section{Introduction}

The good agreement of the observed astrophysical abundances of the 
light elements with those calculated in standard Big-Bang nucleosynthesis (SBBN)~\cite{cfo1} -\cite{cfo5}
imposes interesting constraints on late-decaying massive particles, 
which are generic features of plausible extensions of the Standard Model, such as supersymmetry.
There is even the possibility that the decays of massive particles could improve the
agreement between calculation and observation, particularly in the case of \li7
~\cite{Reno} -\cite{Vasquez:2012dz},
whose abundance is calculated in standard BBN to be considerably higher than that observed
in field halo stars or even globular clusters \cite{cfo5}. In a previous paper~\cite{CEFLOS1.5}, we studied this
possibility in the context of simple supersymmetric models with late-decaying massive
gravitinos, taking into account the uncertainties in the astrophysical observations~\footnote{We
also took into account the uncertainties in the relevant nuclear reaction rates, finding them to be subdominant, which is also the case here.}.
We found that, for suitable values of the gravitino mass $m_{3/2}$ and abundance $\zeta_{3/2} \equiv m_{3/2} n_{3/2}/n_\gamma$,
gravitino decays could reduce the combined $\chi^2$ likelihood function for the deuterium, \he4 and \li7
abundances from the SBBN value of $\sim 32$ to $\sim 5.5$, thus offering a possible solution to the cosmological \li7 problem.

Our previous analysis was in the context of the minimal supersymmetric extension of the
Standard Model with soft supersymmetry-breaking mass parameters constrained to be
universal at the GUT scale (the CMSSM). Moreover, we chose specific values of the
CMSSM parameters that were allowed, even favoured, before the advent of
the LHC searches for supersymmetry. These early parameter choices are now excluded by
both the unsuccessful searches for supersymmetry at the LHC \cite{lhc} and the determination of the Higgs mass at 125-126 GeV \cite{H}. Partly for
this reason, there is increased interest in alternatives to the CMSSM.

In this paper, we revisit our previous study including the WMAP9 estimate of the
baryon density~\cite{wmap} and, where appropriate, updated
estimates of the observational and nuclear-reaction uncertainties. In the absence
of effects due to gravitino decays, we find that the SBBN value of the combined likelihood function
$\chi^2 = 33.7$. Here we study the effects of gravitino decays in CMSSM
models that are favoured in a recent global analysis of the CMSSM parameter space~\cite{MC8},
and extend our study to include models with non-universal Higgs masses~\cite{nuhm1,ELOS} (NUHM1)
and models with universality imposed below the GUT scale~\cite{subGUT,ELOS} (subGUT). 

In all cases, we find
strips in the $(m_{3/2}, \zeta_{3/2})$ plane where the combined $\chi^2$ likelihood function
may be reduced to $\lappeq 3$ if the halo-star estimate of the \li7 abundance is used,
or $< 1$ if the globular-cluster estimate is used (to be compared with an SBBN value of $\sim 23.8$ if globular cluster \li7 is used). If we relax the uncertainty
in the D/H abundance (as may be warranted by the large dispersion in the data), we find that $\chi^2 \sim 1.2$.
We conclude that the late decays
of massive gravitinos offer a robust solution to the cosmological \li7 problem. The likelihood
function for the light-element abundances is minimized for a 
gravitino mass $m_{3/2}$  between $\sim 4.6$ and $\sim 6.2$~TeV and an abundance
$\zeta_{3/2}$ in a narrow range between $\sim 1.0$ and $\sim 2.6 \times 10^{-10}$~GeV in the cases we study.

\section{The Observed Light-Element Abundances and their Uncertainties}

The ranges we adopt for the observed light-element abundances,
follow largely those used in our previous work~\cite{CEFLOS2}. In that paper
we delineated ranges that we considered acceptable, problematic, excluded and
strongly excluded, as specified in Table~\ref{tab:abundances}, which is taken
directly from~\cite{CEFLOS2}. In the subsequent figures, these ranges are left 
unshaded and coloured yellow, red and magenta, respectively. We now
comment on the ranges of the light-element abundances that we consider to be
preferred and use in the subsequent $\chi^2$ analysis: more discussion can be found in~\cite{CEFLOS2}. 

\begin{table}[htb]
\begin{center}
\vspace{0.5cm}
\begin{tabular}{|c||c|c|c|c|c|c|}
\hline\hline
Comparison with & D/H & \he3/D & \he4 & \li6/\li7 & \li7/H & \be9/H \\
observation & $(\times 10^{-5})$ &  &  &  & $\times 10^{-10}$ & $\times 10^{-13}$ \\
\hline\hline
{Strongly excluded} & $< 0.5$ &  $-$ & $< 0.22$ & $-$ & $< 0.1$ & $-$ \\
{Excluded} & $< 1.0$ &  $-$ & $< 0.23$ & $-$ & $< 0.2$ & $-$ \\
{Problematic} & $< 2.3$ &  $-$ & $< 0.24$ & $-$ & $< 0.5 $ & $-$ \\
\hline
Acceptable & [2.3, 3.7] &  [0.3, 1.0] & [0.24, 0.27] & ${< 0.05}$ & [0.5, 2.75] & ${< 0.3}$ \\
\hline
{Problematic} & $> 3.7$ &  $> 1.0$ & $> 0.27$ & $> 0.05$ & $> 2.75$ & $> 0.3$ \\
{Excluded} & $> 5.0$ &  $> 3.0$ & $> 0.28$ & $> 0.1$ & $> 10$ & $> 1.0$ \\
{Strongly excluded} & $> 10$ &  $> 5.0$ & $> 0.29$ & $> 0.2$ & $> 30$ & $> 3.0$ \\
\hline\hline
\end{tabular}
\end{center}
\caption{\it The ranges of light-element abundances whose
comparisons with observation we
consider, following~\cite{CEFLOS2}, to be acceptable, problematic and (strongly) excluded, as shown
in the unshaded/yellow/red/magenta regions in the Figures.}
\label{tab:abundances}
\end{table}

\vspace{0.5cm}
\noindent
\underline{D/H} \\
We assume the central value~\cite{opvs}
\beq
\label{eq:D}
\left( \frac{\rm D}{\rm H} \right)_p = ( 3.01 \pm 0.21 )  \times10^{-5} \, ,
\eeq
corresponding to the deuterium abundance measured in 11 quasar absorption systems~\cite{deut}
and the error in the weighted mean.
The SBBN value for D/H at the current baryon-to-photon ratio is $(2.51 \pm 0.17) \times 10^{-5}$.
The combined uncertainty (theory and observation), is $\pm 0.27 \times 10^{-5}$. We also consider
the sample variance in the data, which leads to a combined uncertainty of $\pm 0.70 \times 10^{-5}$ that is considerably less restrictive. Using the
smaller error corresponds to being more conservative in claiming a solution to the \li7 problem,
whereas the more relaxed version of the D/H constraint corresponds to a more conservative
approach to excluding scenarios. We note that we have
taken a conservative approach in that 
we have not included the recent measurement 
of D/H in \cite{pc}, as the spectrum suffers from the lack of a visible Lyman $\alpha$ emission feature
since the absorber and emitter are at the same redshift. Inclusion of this point
would bring the mean downward (to $2.65 \times 10^{-5}$) with (in our view)
an unrealistically small uncertainty of $0.11 \times 10^{-5}$ 
(and sample variance of $0.36 \times 10^{-5}$). Furthermore,
until several such measurements confirm the precision of this value, we
cannot be assured that no {\it in situ} destruction of D/H has taken place, given the dispersion
in the extant data \cite{opvs}. 

\vspace{0.5cm}
\noindent
\underline{\he3/D} \\
As discussed in~\cite{CEFLOS2}, we expect the ratio
\he3/D \cite{sigl} to be a monotonically increasing function of time, and therefore consider
the solar ratio \he3/D $\simeq 1$~\cite{geiss} to be a conservative upper bound on the BBN ratio. As we
see later, this constraint is satisfied comfortably within all the region of interest, so we do not
include it in our global $\chi^2$ function, and hence do not need to assign an uncertainty.

\vspace{0.5cm}
\noindent
\underline{\he4} \\
Following~\cite{CEFLOS2}, we use the estimate~\cite{aos3} 
\beq
Y_p = 0.2534 \pm 0.0083
\eeq
based on an analysis of the regression of $Y$ vs. O/H.
This should be compared to the SBBN value of 0.2487 $\pm$ 0.0002.

\vspace{0.5cm}
\noindent
\underline{\li6/\li7} \\
As in~\cite{CEFLOS2}, we assign an upper limit $\li6/\li7 < 0.05$, and (like the \he3/D constraint)
do not include it in our global $\chi^2$ function.

\vspace{0.5cm}
\noindent
\underline{\li7/H} \\
The cosmological \li7 problem \cite{cfo5} refers to the discrepancy between the 
SBBN prediction of \li7/H  $= (5.11^{+0.71}_{-0.62}) \times 10^{-10}$ and observations~\cite{li7obs}. 
We adopt here the observational range \cite{rbofn}
\begin{equation}
\left( \frac{\li7}{\rm H} \right)_{\rm halo \star} \; = \; (1.23^{+0.34}_{-0.16}) \times 10^{-10} ,
\label{Li7Hhalo}
\end{equation}
leading to a combined uncertainty that we take as $\pm 0.71 \times 10^{-10}$.
We note, however, that the \li7 abundance in globular cluster stars is reported to be
a factor $\sim 2$ higher \cite{liglob,liglob2}. In our analysis below,
we also derive results assuming \cite{liglob} \li7/H $= (2.34 \pm 0.05) \times 10^{-10}$
with a combined uncertainty of $\pm 0.62 \times 10^{-10}$.

\vspace{0.5cm}
\noindent
\underline{\be9/H} \\
We note finally that \be9 is also observed in halo dwarf stars, with an abundance that
varies strongly with metallicity, see, e.g.,~\cite{boes}.
The observation with the lowest metallicity has an [O/H] value of about -2.5 with a \be9/H abundance of 
$3 \times 10^{-14}$. We consider this to be a conservative upper limit on the \be9/H abundance,
while noting that there is one observation \cite{ito} of \be9 with an abundance about 3 times lower.
As in the cases of \he3/D and \li6/\li7, we do not include this observable in our global $\chi^2$ function.

Finally, we note that our SBBN results are based on the WMAP9 determination of the 
baryon density~\cite{wmap} that corresponds to a baryon-to-photon ratio of $\eta = (6.20 \pm 0.14) \times 10^{-10}$.
WMAP9 also provides a value of the cold dark matter density that can be used as an upper limit
on the possible cosmological density of supersymmetric dark matter neutralino particles $\chi$ produced through thermal freeze out and by
gravitino decays: $\Omega_\chi h^2 \le 0.1138 \pm 0.0045$.
In all of the models considered below, the relic density of neutralinos
from thermal annihilations has been adjusted to yield the central
WMAP value of $\Omega_\chi h^2 = 0.1138$ using the {\tt SSARD} code~\cite{ssard}, 
so all the neutralinos produced by gravitino decays contribute to the global
$\chi^2$ we define below.

\section{Supersymmetric Models Studied}

We have studied the light-element abundances in ten different supersymmetric models
that respect the LHC and other current constraints.

Four of these models are taken from~\cite{MC8}~\footnote{Their parameters are adapted for study using the
{\tt SSARD} code~\cite{ssard}.}, which studied the parameter spaces of the CMSSM
and the NUHM1, taking into account not only the direct searches for supersymmetry via
missing-transverse-energy events at the LHC~\cite{lhc}, but also the mass measured for
the Higgs boson discovered by ATLAS and CMS~\cite{H} and searches 
for the rare decay $B_s \to \mu^+ \mu^-$~\cite{bmm}~\footnote{The results in~\cite{MC8} are very little
affected by incorporation of the recent measurement of \ensuremath{{\rm BR}(B_s \to \mu^+\mu^-)} by the 
LHCb Collaboration~\cite{LHCbnew},
whose central value is quite close to the combination of previous data, and very consistent with the
predictions of the CMSSM and NUHM1 fits in~\cite{MC8}:
see {\tt http://mastercode.web.cern.ch/mastercode/} for this and other current results from the Mastercode group.
The best-fit models of~\cite{MC8} are also compatible with the preliminary constraints from
LHC 8-TeV data~\cite{LHC8}.}
and the XENON100 direct search for dark matter scattering~\cite{XE100}. Ref.~\cite{MC8} found two well-defined and almost equally good local
minima of the global likelihood function for the CMSSM, with quite distinct values of the scalar
and gaugino soft supersymmetry-breaking parameters, $m_0$ and $m_{1/2}$, and the ratio of Higgs
v.e.v.s, $\tan \beta$. Their values at the local best fits are shown in the first few columns of the first
two rows in Table~\ref{tab:bestfits}. The first of these points lies on a strip where the density of 
the lightest supersymmetric particle (LSP), the neutralino $\chi$, is brought into the WMAP range
by coannihilations with the nearly-degenerate next-to-lightest sparticle, the lighter stau  ${\tilde \tau_1}$.
The second point is located in a rapid-annihilation funnel where s-channel annihilations
through heavy Higgses control the relic density. We also list the corresponding best-fit values for 
the trilinear soft supersymmetry-breaking parameter, $A_0$, at the GUT scale, and note the positive sign assumed in~\cite{MC8}
for the Higgs mixing parameter, $\mu$, whose numerical value is fixed by the electroweak vacuum conditions in the CMSSM. The next two rows of Table~\ref{tab:bestfits} list the corresponding parameters for the best fit adapted from~\cite{MC8}
in the NUHM1 framework, and a representative good fit with larger values of $m_0, m_{1/2}$ and $\tan \beta$. In
these cases the corresponding values of $\mu$ are not fixed by the electroweak vacuum conditions, and we note their
fit values.

\begin{landscape}
\thispagestyle{empty}
\begin{table*}[!tbh!]
\renewcommand{\arraystretch}{1.5}
{
\begin{tabular}{|c|c|c||c|c|c|c|c||c|c||c|c|c|c|} \hline
ID & Model & Ref & $m_{1/2}$ & $m_{0}$ & $A_0$ & $\tan \beta$ & $\mu$ & $m_\chi$ & $m_h$ 
  & $ {m}_{3/2}$ & ${\zeta}_{3/2}$ & $\tau_{3/2}$ & $\chi^2_{\rm min}$ \\
\hline \hline
1 & CMSSM & \cite{MC8}
    &   $905$ & $361$ 
    & $1800$ & $16$ & $> 0$ & 395 & 123.8 
    & 4560 & $1.5 \times 10^{-10}$ & 208 & 2.81 \\
2 & CMSSM & \cite{MC8}
    & $1895$ & $1200 $ 
    & $1200$ & $50$  &  $> 0$ & 857 &123.3 
    & 5520 & $1.8 \times 10^{-10}$ & 231 &  2.86 \\
3 & NUHM1 & \cite{MC8}   & $970$ & 345     & 2600 & $15$ & 2600 & 427 & 123.8 
    & 4600 & $1.2 \times 10^{-10}$ & 220 & 2.82 \\
4 & NUHM1 & \cite{MC8}   & $2800$ & 1040     & 2100 & $39$ &  3800 & 1288 &124.0 
    & 6200 & $2.6 \times 10^{-10}$ & 276 & 3.14 \\
\hline
5 & CMSSM & Fig.~2d of \cite{ELOS}
    & $1115$ & $1000$ & 2500 & 40 & $> 0$ & 496 & 124.8 
    & 4800 & $1.6 \times 10^{-10}$ & 213 & 2.87 \\
6 &   NUHM1 & Fig.~5b of \cite{ELOS}  
    & $1175$ & $1500$ & 3000 & 40 &  500  & 499 & 125.9 
    & 5000 & $2.6 \times 10^{-10}$ & 188 & 2.86 \\
7 &   NUHM1 & Fig.~6b of \cite{ELOS}  
    & $1300$ & $1000$ & 2500 & 30 & -550 & 550 & 125.5 
    & 4700 & $1.0 \times 10^{-10}$ & 258 & 2.87 \\
8 &    subGUT CMSSM & Fig.~9c of \cite{ELOS} 
    & $2040$ & $2200$ & 5500 & 10 & $> 0$ & 1554 & 126.7 
    & 5400 & $1.6 \times 10^{-10}$ & 214 & 2.96 \\
9 &    subGUT mSUGRA & Fig.~10d of \cite{ELOS}
    & 2400 & 4000 & Polonyi & 36 & $> 0$ & 1099 & 125.4 
    & 6000 & $1.6 \times 10^{-10}$ & 239 & 2.91 \\
10 &    subGUT mSUGRA & Fig.~10d of \cite{ELOS}
    & $1700$ & $2000$ & Polonyi & 33 & $> 0$ & 1110 & 124.0 
    & 5100 & $1.6 \times 10^{-10}$ & 219 & 2.89 \\
\hline
11 & CMSSM${}^{\rm (a)}$ & \cite{MC8}
    &   $905$ & $361$ 
    & $1800$ & $16$ & $> 0$ & 395 & 123.8 
    & 4440 & $1.5 \times 10^{-10}$ & 230 & 1.25 \\
12 & CMSSM${}^{\rm (b)}$ & \cite{MC8}
    &   $905$ & $361$ 
    & $1800$ & $16$ & $> 0$ & 395 & 123.8 
    & 4520 & $1.0 \times 10^{-10}$ & 215 & 0.52 \\
13 & CMSSM${}^{\rm (c)}$ & \cite{MC8}
    &   $905$ & $361$ 
    & $1800$ & $16$ & $> 0$ & 395 & 123.8 
    & 4360 & $7.1 \times 10^{-11}$ & 245 & 0.37 \\
\hline
\end{tabular}
 }
\caption{\it The models studied, with references, their input parameters and the corresponding values of
$m_h$ calculated using {\tt FeynHiggs}~\protect\cite{FH} (which have an estimated theoretical uncertainty $\gappeq 1.5$~GeV),
the best-fit gravitino mass $m_{3/2}$ and abundance $\zeta_{3/2}$, and the minimum $\chi^2$ in a global
fit to the observed values of the light-element abundances. All mass parameters are expressed in GeV units, and the best-fit lifetime $\tau_{3/2}$ is in seconds.
The subGUT model in Fig.~9c of~\protect\cite{ELOS} assumes $M_{in} = 10^9$~GeV, and those in Fig.~10d
of~\protect\cite{ELOS} assume $M_{in} = 10^{10}$~GeV.
In models 11 through 13,
the $\chi^2$ and best-fit values are computed using (a) 
the D/H sample variance uncertainty, (b) the \li7/H abundance as 
determined from globular clusters, and (c)
both the D/H sample variance uncertainty and the globular
cluster \li7/H.
 }
\label{tab:bestfits}
\end{table*}
\end{landscape}

We also show in Table~\ref{tab:bestfits} the corresponding masses of the lightest
neutralino, $\chi$, which is assumed in this paper to be the LSP and to
constitute (the bulk of) the dark matter. We note that $m_\chi$ varies over a large range in the model fits.
We also note the values calculated for the mass of the Higgs boson,
$m_h$. Considering the theoretical uncertainties in the calculation of $m_h$ \cite{FH}, these values are
consistent with the LHC measurement $m_h \sim 125$ to 126~GeV~\cite{H}.

Although the first four rows in Table~\ref{tab:bestfits} include the best fits in the CMSSM and NUHM1,
they do not exhaust the interesting possibilities offered by these models. Accordingly, we have also
studied three other choices of CMMSM and NUHM1 parameters that are also compatible with the
LHC constraints and offer complementary
possibilities. The fifth row lists the parameters of a point towards the tip of the $\chi - {\tilde \tau_1}$ coannihilation
strip for $\tan \beta = 40$ shown in Fig.~2d of~\cite{ELOS}, which has larger values of $m_0, m_{1/2}, A_0$ and
$m_\chi$ than the best-fit point in the first row of Table~\ref{tab:bestfits}. The sixth row illustrates a new
possibility in the NUHM1 compared to the CMSSM shown in Fig.~5b of~\cite{ELOS}, namely the appearance at 
large $m_{1/2}$ of a band where the relic LSP density is brought down into the cosmological range because the 
make-up of the $\chi$ is in transition between a gaugino-like state at lower $m_{1/2}$ which has
too high a relic density, and a Higgsino-like state at higher $m_{1/2}$, which has too low a relic density.
The seventh row in Table~\ref{tab:bestfits} is an NUHM1 example with similar values of $m_0, m_{1/2}$ and 
$\tan \beta$, but with $\mu < 0$, which also lies along a transition strip, as seen in Fig.~6b of~\cite{ELOS}.

Finally, the next three rows in Table~\ref{tab:bestfits} illustrate possibilities if the soft supersymmetry-breaking
parameters are universal at some scale $M_{in} < M_{GUT}$. The eighth row is a CMSSM example with
$M_{in} = 10^9$~GeV, which is located along one of the rapid-annihilation funnels visible in Fig.~9c 
of~\cite{ELOS}. The next two are mSUGRA-inspired examples with 
$M_{in} = 10^{10}$~GeV that are located on either side of the prominent rapid-annihilation
funnel visible in Fig.~10d of~\cite{ELOS}. In these cases supersymmetry is assumed 
to be broken by the simplest Polonyi superpotential, which specifies the corresponding value of $A_0$.
These examples populate regions of the $(m_{1/2}, m_0)$ plane that are not allowed in the CMSSM or
the NUHM1 examples given above: they all have $m_{1/2} < m_0$ and large values of $m_\chi$.

Since these examples exhibit several features not present in the specific CMSSM models studied
previously, they serve to probe the robustness of the scenario for resolving the cosmological \li7
problem proposed in~\cite{CEFLOS, CEFLOS1.5}.

\section{Results}

In this Section we present in some detail results for the LHC-favoured CMSSM and NUHM1
models presented in the first four rows of Table~\ref{tab:bestfits}, and then summarize results
for the other six models.

In Fig.~\ref{fig:zeta_m32} we display the effects on the light-element abundances 
of the late decays of a gravitino with mass $m_{3/2} \in [2, 8]$~TeV and
abundance $\zeta_{3/2} \in [10^{-14}, 10^{-7}]$~GeV. We show blocks of six panels for
each of the models described in the first four rows of Table~\ref{tab:bestfits}: lower-mass CMSSM
(upper left), higher-mass CMSSM (upper right), lower-mass NUHM1 (lower left) and higher-mass
NUHM1 (lower right). For each model, the upper
left panel displays the effect of gravitino decays on the D/H ratio, the upper middle panel the effect on the \he3/D ratio, the upper
right panel the effect on the \he4 abundance, the lower left the effect on the \li6/\li7 ratio, the lower middle the effect on the
\li7/H ratio, and the lower right the effect on the \be9/H ratio. In each panel, the unshaded, yellow, red and magenta regions correspond to the ranges specified in Table~\ref{tab:abundances}. 
Solid shadings are used for regions with excess abundances, and hashed shadings for regions with low abundances. 

The corresponding panels for the different models exhibit several common features.
We note first that the D/H constraint typically excludes a large triangular region at
small $m_{3/2}$ and large $\zeta_{3/2}$, but has no impact in the triangular region
at large $m_{3/2}$ and small $\zeta_{3/2}$. Next we note that the \he3/D constraint
has no impact in any of these models. The impact of the \he4 constraint is always
limited to a small region at relatively small $m_{3/2}$ and large $\zeta_{3/2}$ that
lies within the region already excluded by the D/H ratio. The \li6/\li7 ratio typically excludes
a band at smaller $m_{3/2}$ that is cut off at very small $\zeta_{3/2}$. The \li7/H
ratio excludes not only a region at small $m_{3/2}$ and large $\zeta_{3/2}$, but also
a region at large $m_{3/2}$ and small $\zeta_{3/2}$. In between, there is a banana-shaped 
region extending from $m_{3/2} \sim 3$~TeV upwards, where the \li7 abundance falls
within acceptable limits. The yellow shading at lower right in each of these
panels reflects the existence of the cosmological \li7 problem in standard BBN, and the
unshaded banana indicates that this may be solved by the late decays of massive
gravitinos. Finally, we note that the \be9/H constraint has impact only at small $m_{3/2}$
and large $\zeta_{3/2}$, in a region already excluded by the D/H ratio.

It is clear from this survey that, in all these four models, the most important constraints
are those from the D/H and \li7/H ratios, and the issue is whether, within the unshaded
\li7 banana, there are model parameters where these constraints can be reconciled
adequately. 

This issue is addressed in Fig.~\ref{fig:D_Li} for the CMSSM and NUHM1 models
described in the first four rows of Table~\ref{tab:bestfits}. As in Fig.~\ref{fig:zeta_m32},
the upper left panel is for the lower-mass CMSSM model, the upper right panel for the higher-mass
CMSSM model, the lower left panel for the lower-mass NUHM1 model, and the lower right
panel for the higher-mass NUHM1 model. In each panel, we display the values of D/H and \li7/H
for a grid of points in the $(m_{3/2}, \zeta_{3/2})$ plane.
The pattern of points in Fig.~\ref{fig:D_Li} arises as follows.  Each of the
loops (best visible at moderate D/H and \li7/H) drops
nearly vertically, goes through a minimum, and then
rises towards high \li7/H and low D/H.  A single loop
represents a single value of $\zeta_{3/2}$.
As seen in Fig.~\ref{fig:zeta_m32}, scanning from low to high $m_{3/2}$ at fixed
$\zeta_{3/2}$ generally takes us from high D/H to
low D/H, and from high \li7/H to low \li7/H to somewhat
higher \li7/H.  This leads to the loop behavior.
For example the top ``row" of blue points seen in the two panels on the left, correspond
to $m_{3/2} = 2$~TeV and $\zeta_{3/2}$ increases from left to right. 
We see that the points examined in this scan grazes the
inner ellipse representing the combined one-$\sigma$ excursion of the observed D/H and
\li7/H ratios. 

The predominant feature in Fig.~\ref{fig:D_Li} is the correlation between
lower \li7/H and higher D/H that is a common feature of many potential
solutions to the \li7 problem \cite{opvs}.  This is in fact a welcome feature
as some enhancement in D/H is warranted when comparing the 
mean value from observations and the SBBN result. 
The red crosses represent the best-fit points in each of the four models, and lie close to
the one-$\sigma$ ellipses in all cases. We learn from these plots that the D/H
and \li7/H observations can indeed be reconciled, with the D/H and \li7/H ratios
lying within their combined one-$\sigma$ observational band, somewhat
above the central values given in Section~2.

\begin{landscape}
\thispagestyle{empty}
\begin{figure}[ht!]
\begin{center}
\epsfig{file=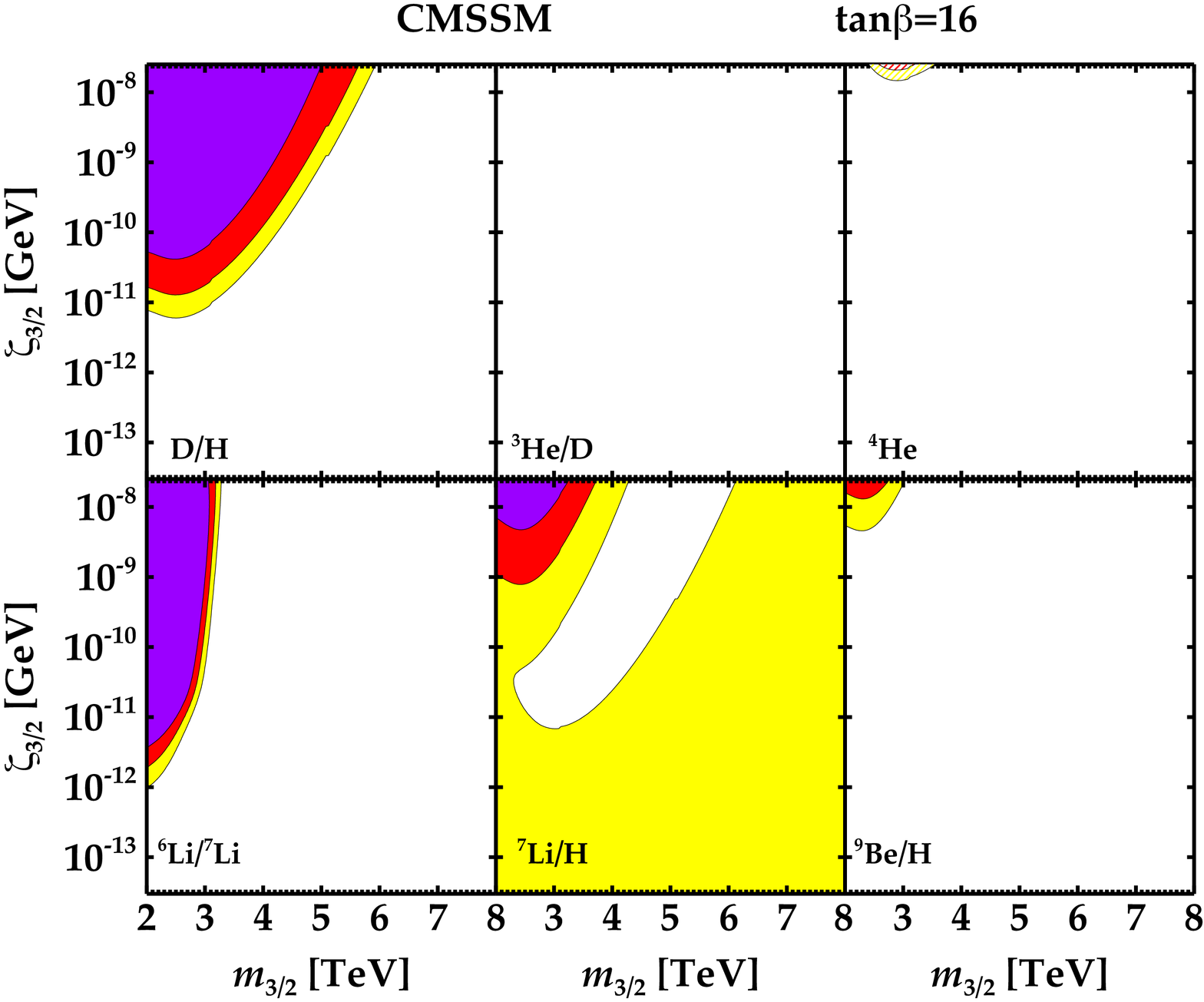,angle=90,height=7cm}
\epsfig{file=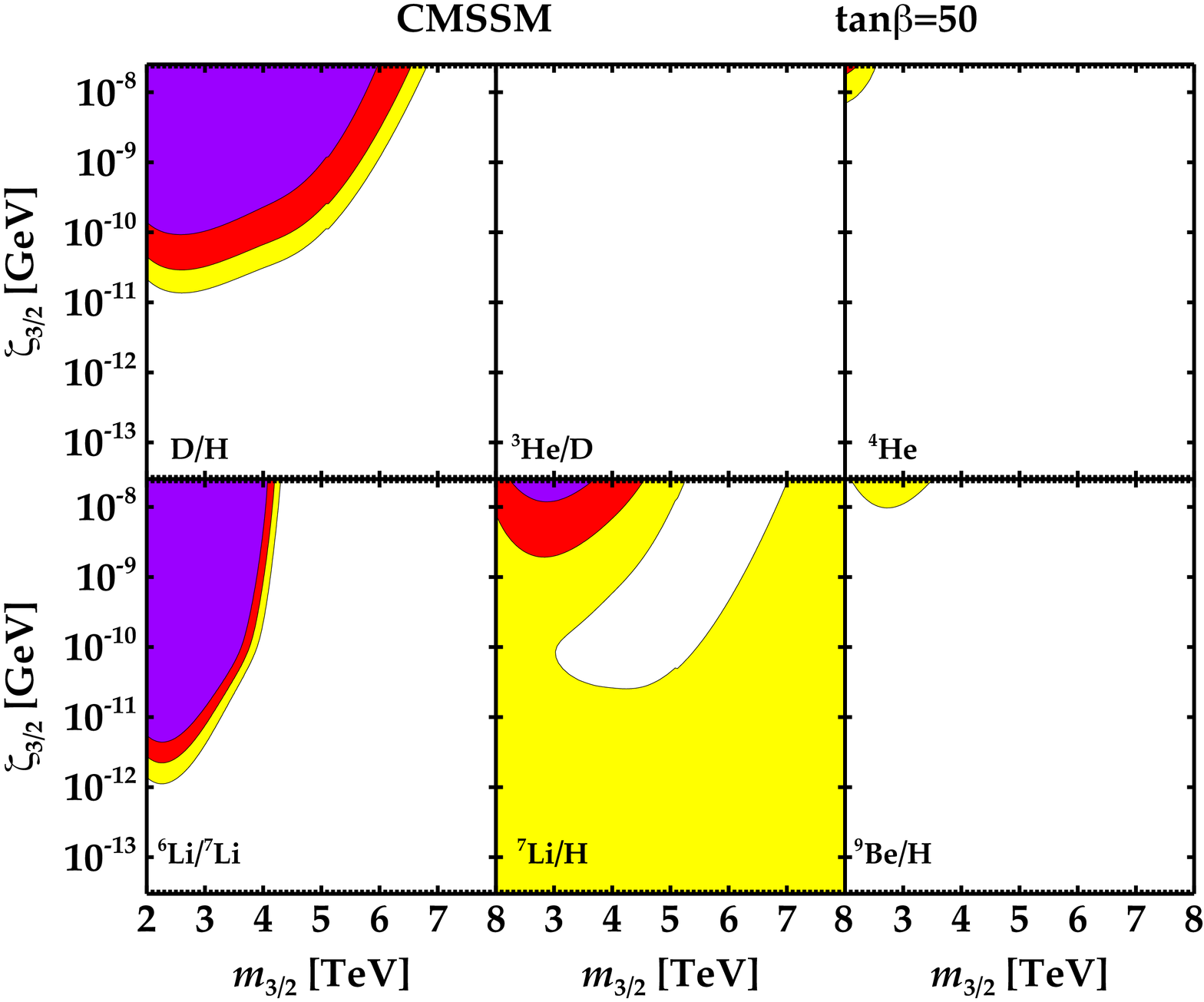,angle=90,height=7cm}\\
\epsfig{file=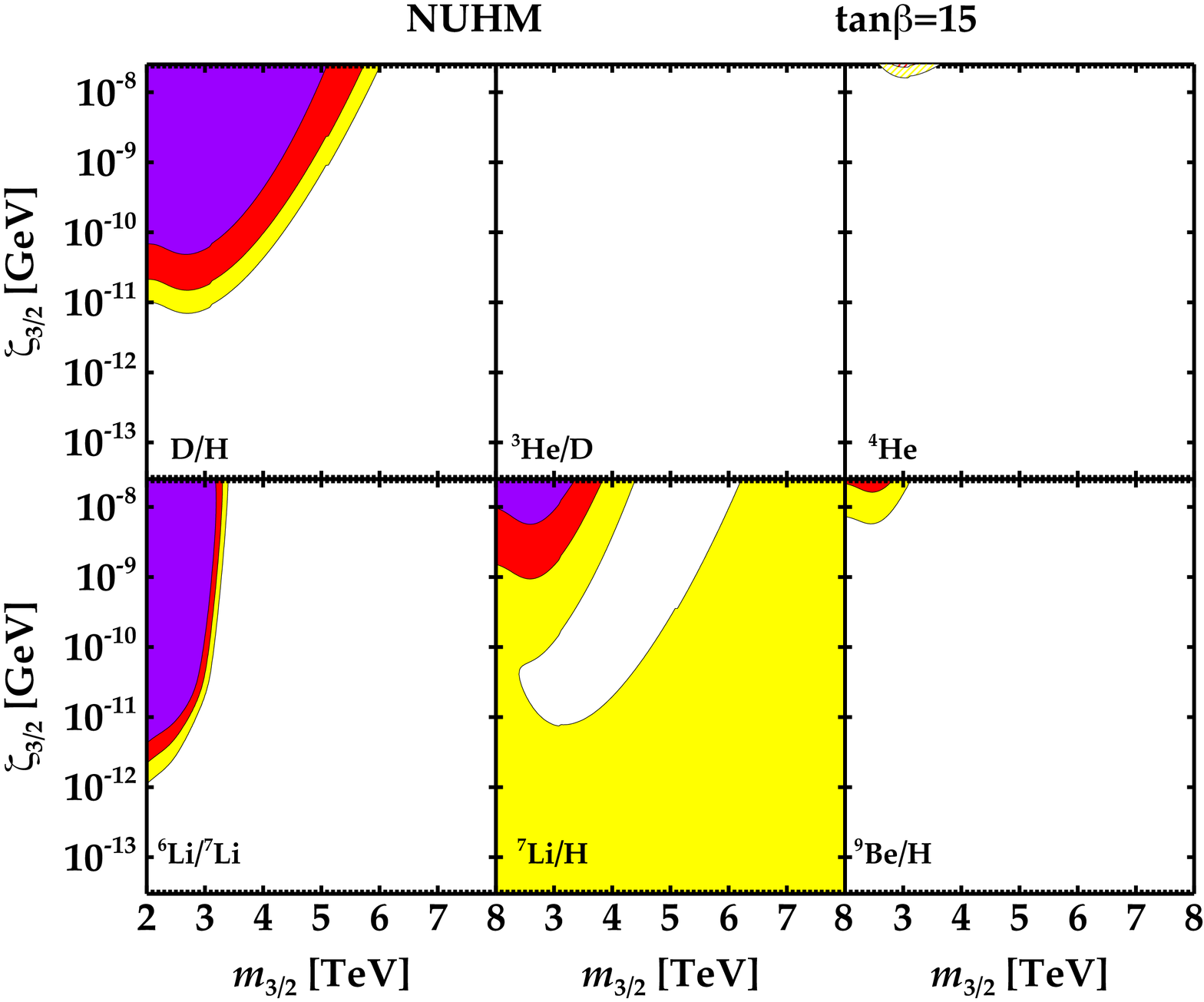,angle=90,height=7cm}
\epsfig{file=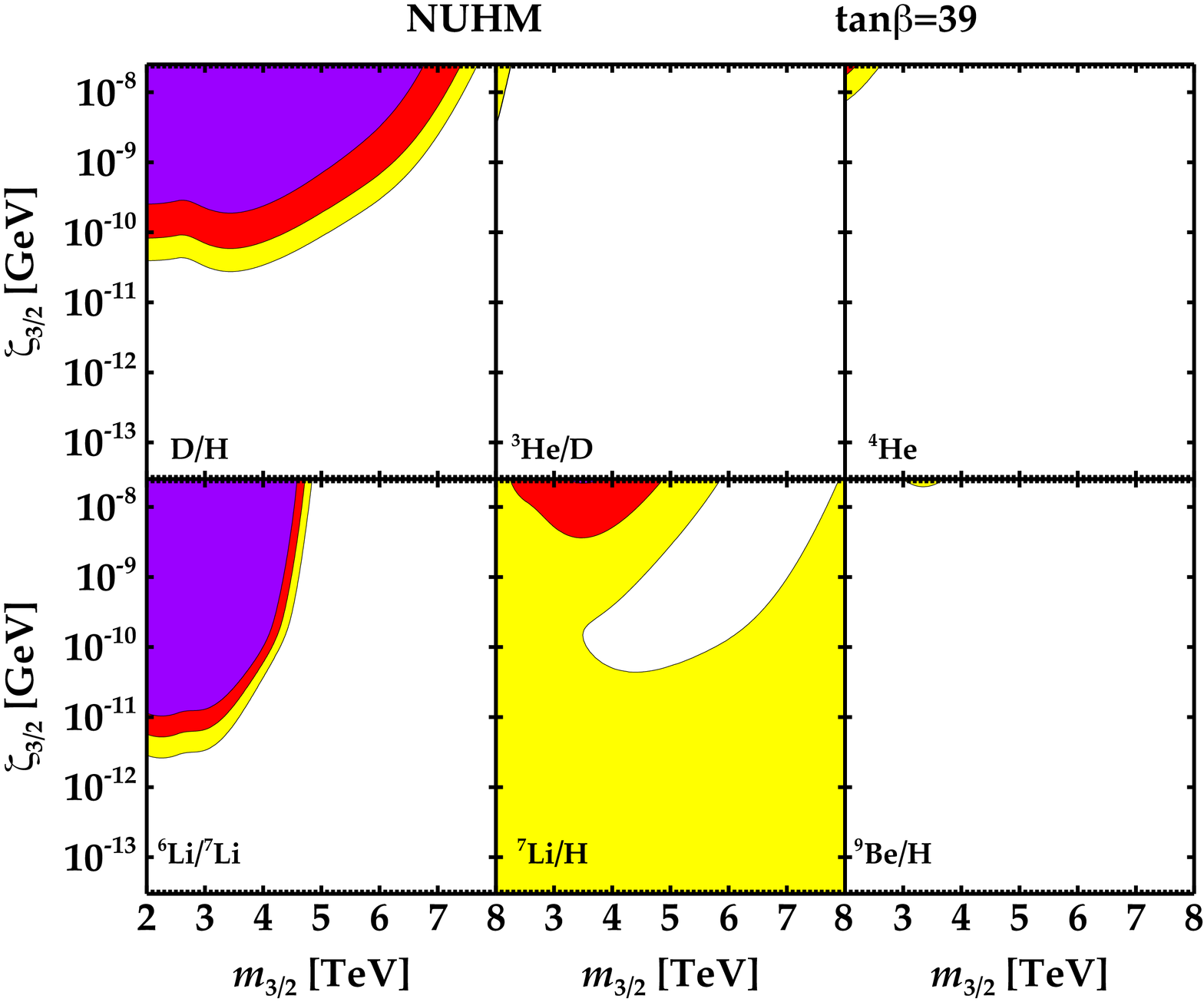,angle=90,height=7cm}
\end{center}
\vspace{-1cm}
\caption{
\it 
Plots of the effects on the light-element abundances 
of the decays of a gravitino with mass $m_{3/2} \in [2, 8]$~TeV and
abundance $\zeta_{3/2} \in [10^{-14}, 10^{-7}]$~GeV. The results are displayed in blocks of six panels for
each of the models described in the first four rows of Table~\ref{tab:bestfits}. In each block, the upper
left panel displays the effect on the D/H ratio, the upper middle panel that on the \he3/D ratio, the upper
right panel that on the \he4 abundance, the lower left that on the \li6/\li7 ratio, the lower middle that on the
\li7/H ratio, and the lower right that on the \be9/H ratio. The unshaded regions in the panels are those allowed
at face value by the ranges of the light-element abundances
reviewed in Section~2,
whilst the yellow, red and magenta regions correspond to
progressively larger deviations from the central values of the abundances,
as summarized in Table~\ref{tab:abundances}. 
Solid shadings are used for regions with excess abundances, and hashed shadings for regions with low abundances. 
\label{fig:zeta_m32}}
\end{figure}
\end{landscape}

\begin{figure}[h!]
\begin{center}
\epsfig{file=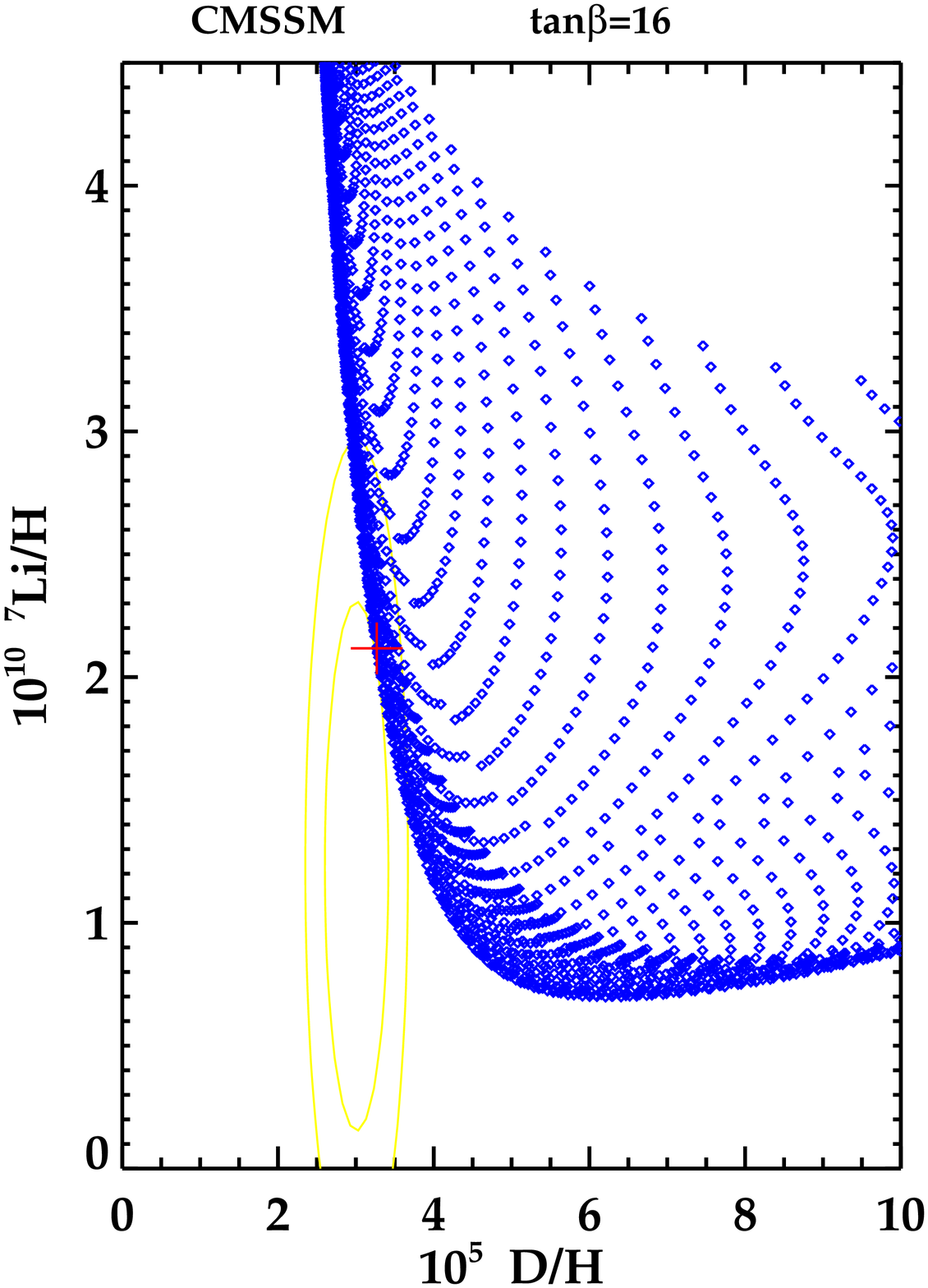,height=8cm}
\epsfig{file=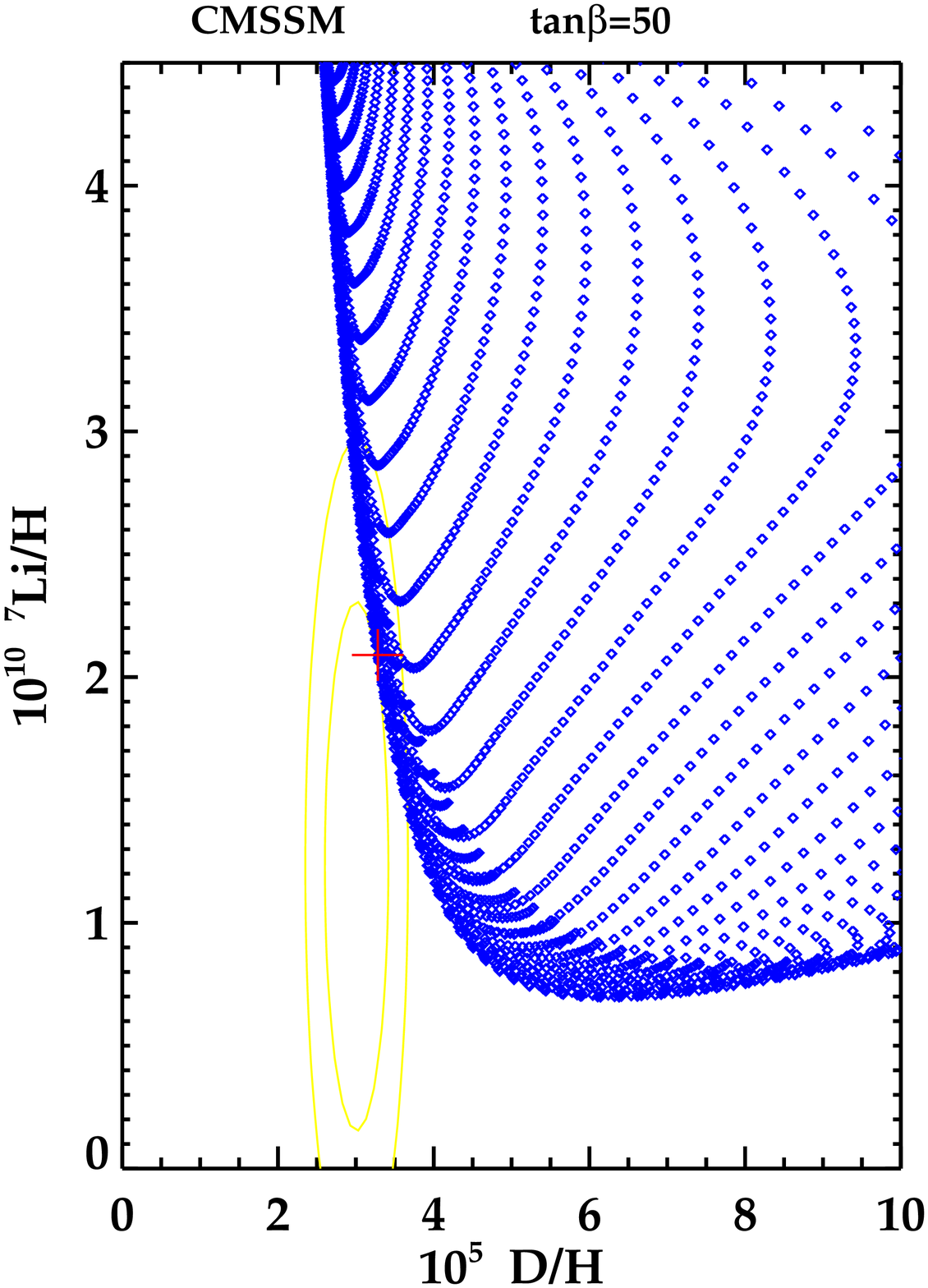,height=8cm}\\
\epsfig{file=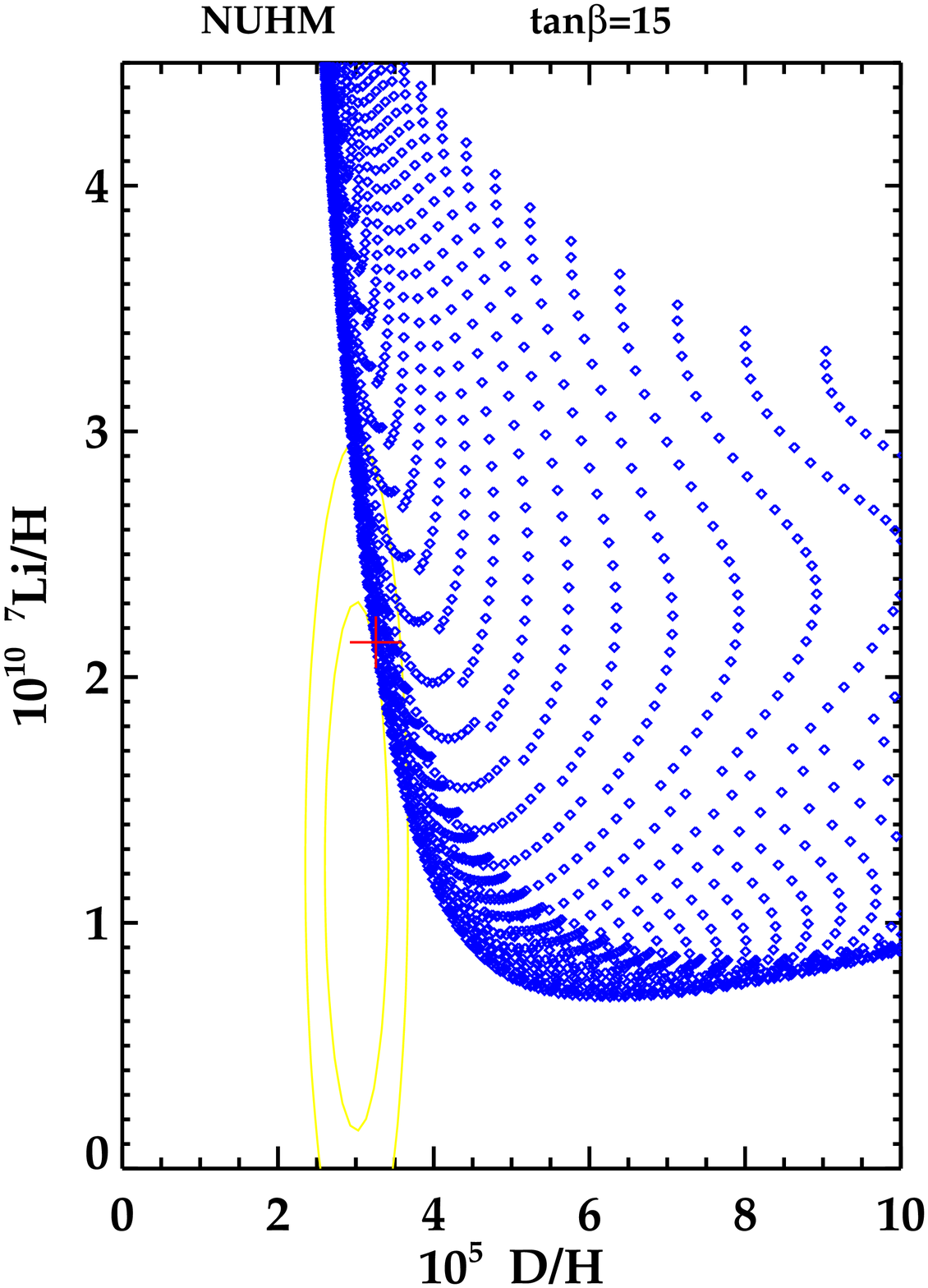,height=8cm}
\epsfig{file=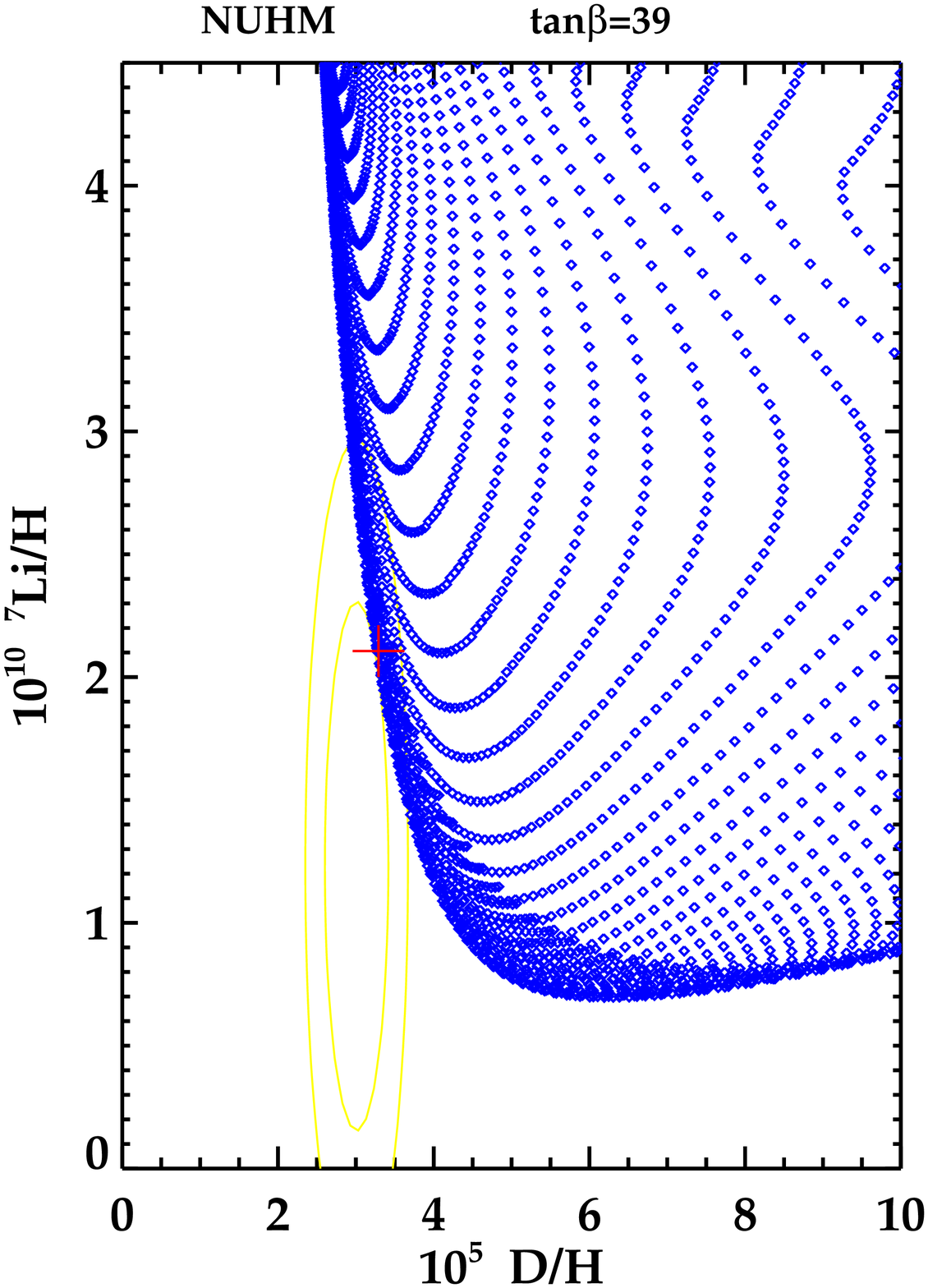,height=8cm}
\end{center}
\caption{\it
Planes giving overviews of the D/H and \li7/H abundances obtainable in the models
described in the first four rows of Table~\ref{tab:bestfits}. The blue points
show the values found in scans of the models for different values of
$m_{3/2}$ and $\zeta_{3/2}$. The ellipses represent the one- and two-$\sigma$
regions found by combining the D and \li7 constraints reviewed in Section~2. The red cross
marks the best fit found in each model.}
\label{fig:D_Li}
\end{figure}

\begin{figure}
\begin{center}
\epsfig{file=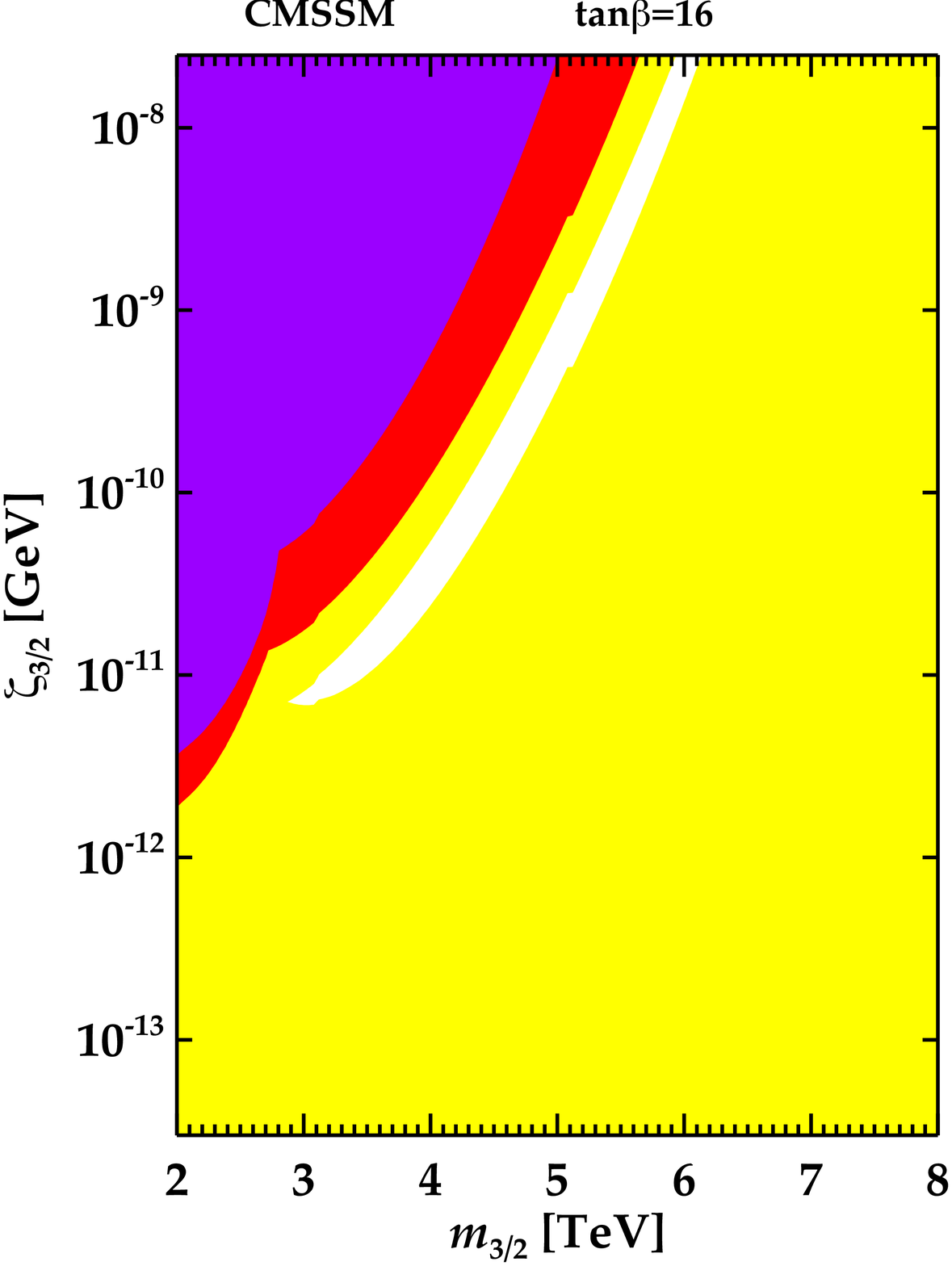,height=8cm}
\epsfig{file=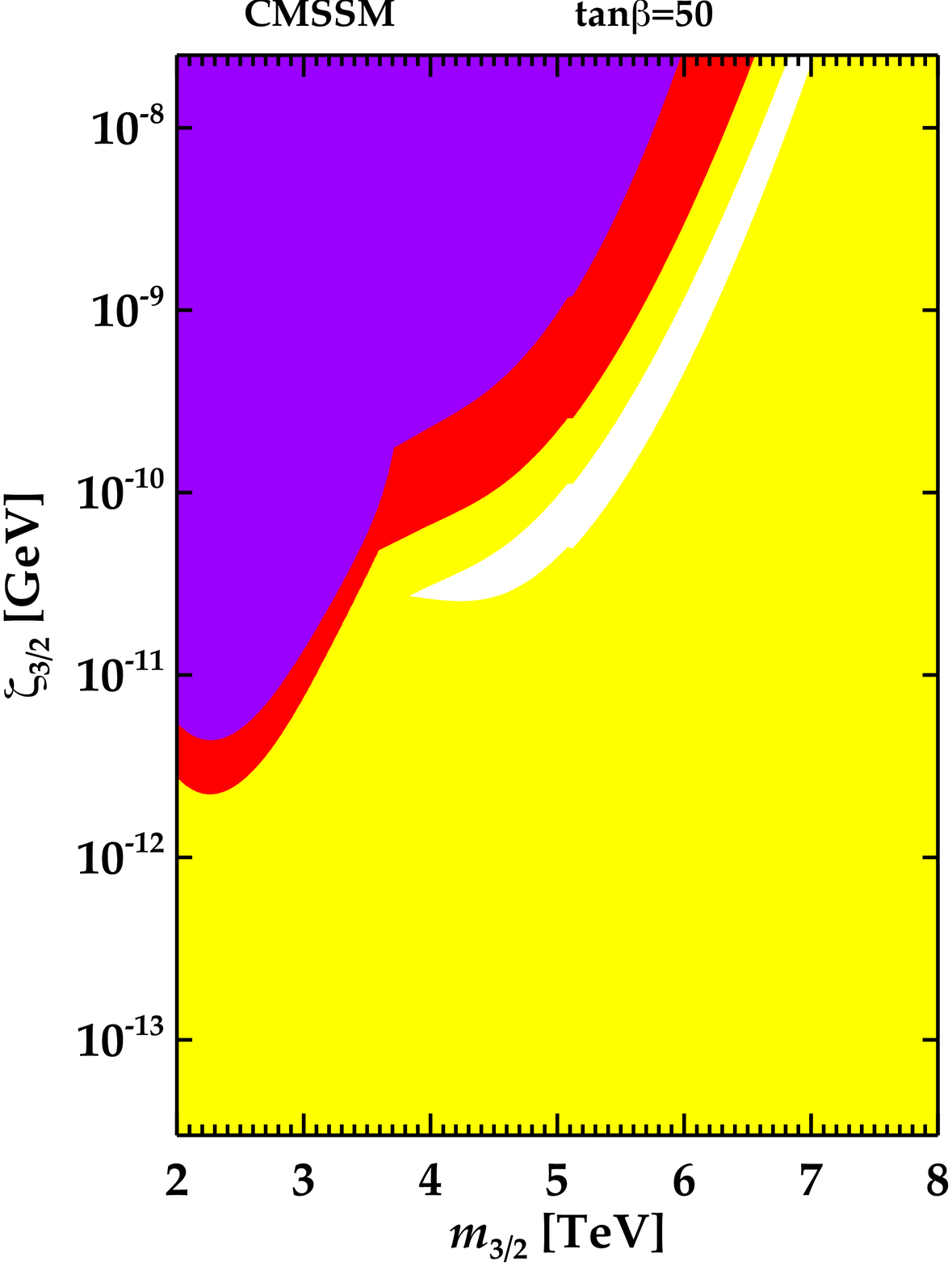,height=8cm}\\
\epsfig{file=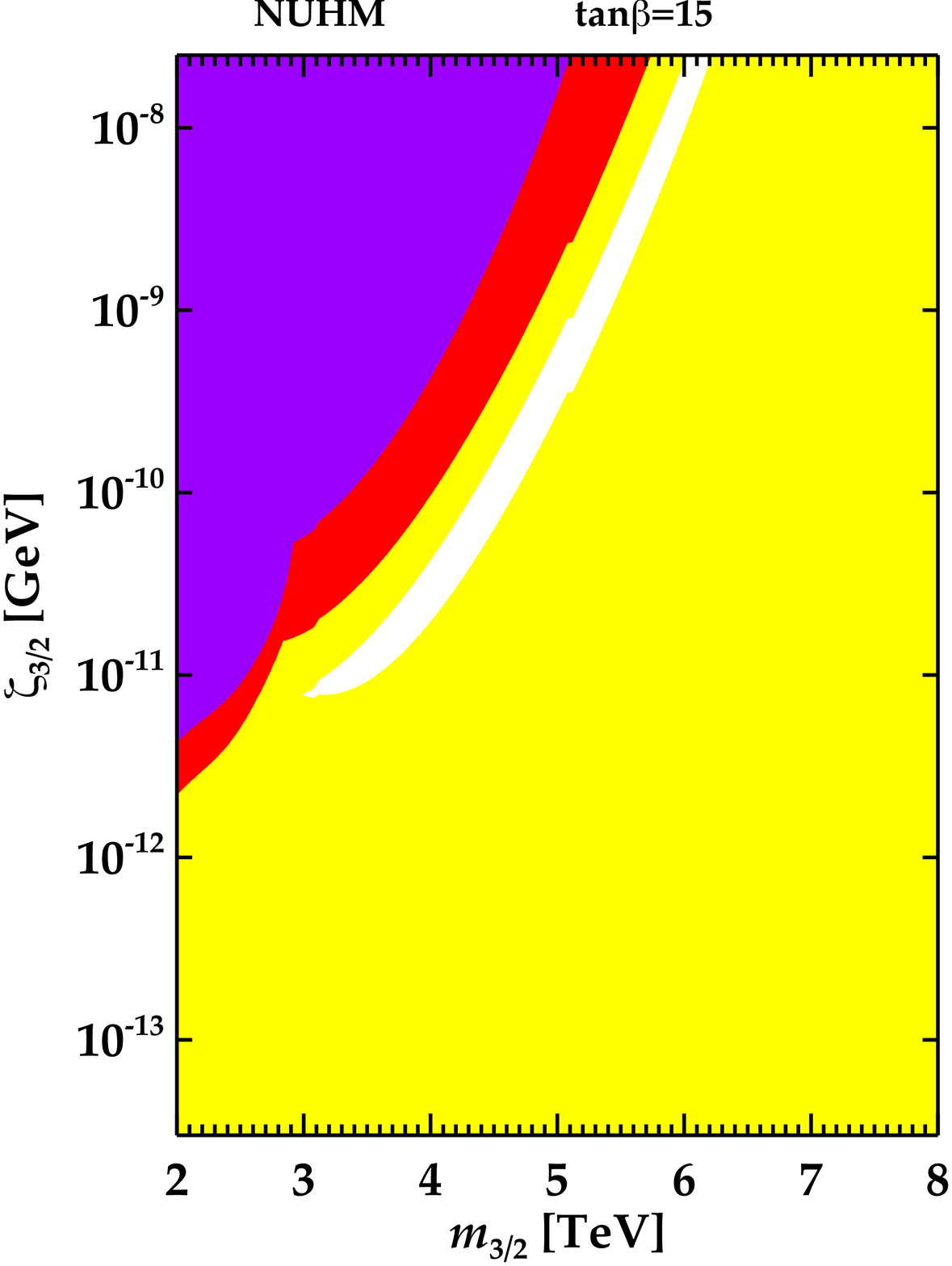,height=8cm}
\epsfig{file=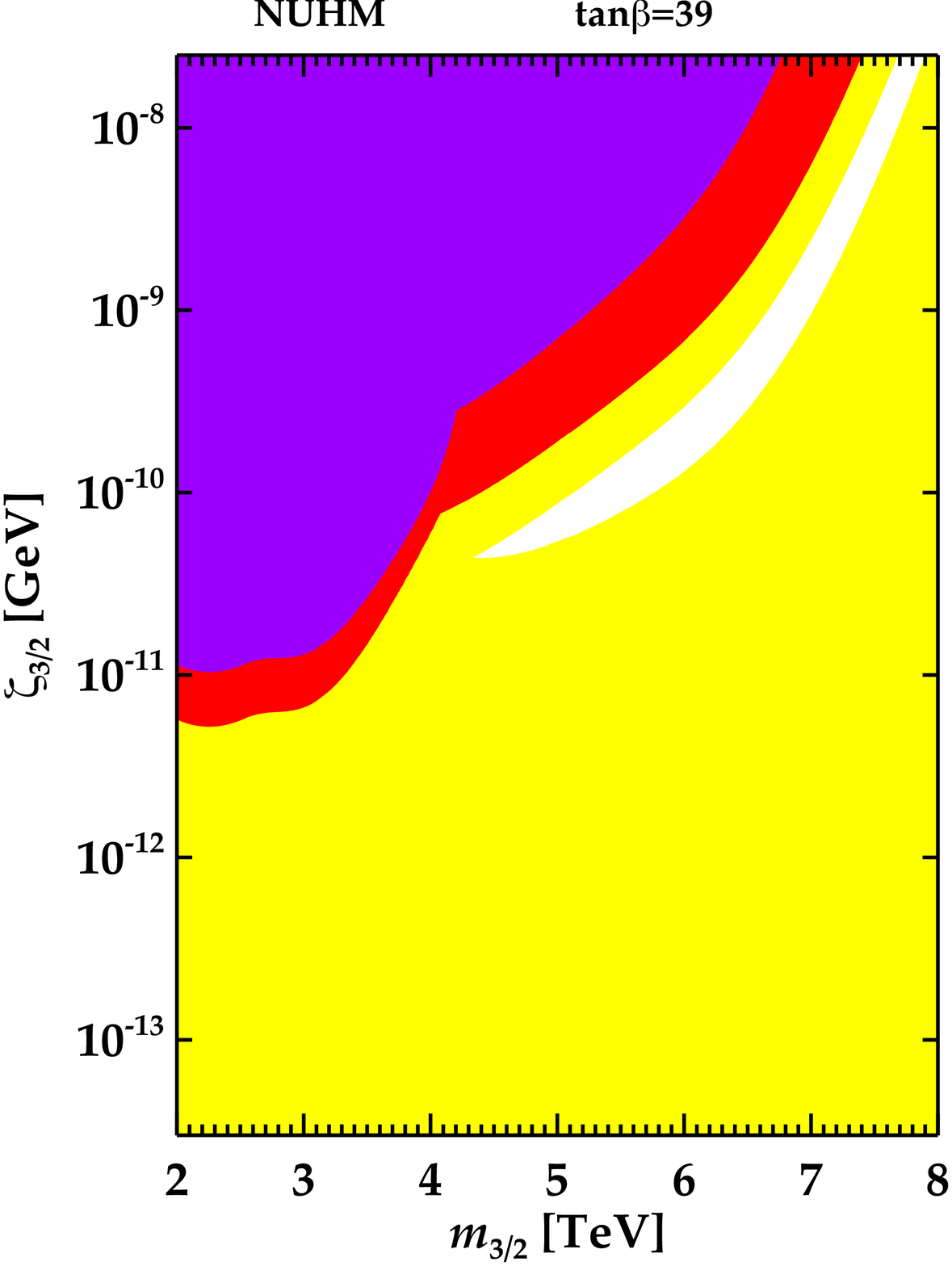,height=8cm}
\end{center}
\caption{\it
Planes exploring the compatibility of all the light-element constraints in the
$(m_{3/2}, \zeta_{3/2})$ planes for the models described in the first four rows of
Table~\ref{tab:bestfits}. In each case, the magenta region is where one or more calculated
abundance lies within a `strongly-excluded' range specified in Table~\ref{tab:abundances},
whereas in the red region the largest deviation is in an `excluded' range, and in the yellow
region the largest deviation is in a `problematic' range. In the unshaded banana-shaped regions, 
there is no significant deviation in any of the calculated abundances.}
\label{fig:summary}
\end{figure}

The broader issue of compatibility with all the constraints from light-element
abundances is addressed in Fig.~\ref{fig:summary}, which explores the 
mutual compatibility of all the light-element constraints in the
$(m_{3/2}, \zeta_{3/2})$ planes for the models described in the first four rows of
Table~\ref{tab:bestfits}. In each case, the magenta region is where one or more calculated
abundance lies within a `strongly-excluded' range specified in Table~\ref{tab:abundances},
whereas in the red region the largest deviation is in an `excluded' range, and in the yellow
region the largest deviation is in a `problematic' range. 
In each panel, we see that there is a strongly excluded magenta region
at smaller $m_{3/2}$ and larger $\zeta_{3/2}$ (where there are deviations for D/H and
some other abundances) and a disfavoured region at 
larger $m_{3/2}$ and smaller $\zeta_{3/2}$ (where the \li7 abundance is similar to that
in standard BBN) separated by an unshaded banana-shaped strip where no calculated
light-element abundance disagrees significantly with observation. While the exact location
and shape of the unshaded banana differs in each model, its presence
is a generic feature we see not only in the four models displayed but in all ten
we have studied.

\begin{figure}
\begin{center}
\epsfig{file=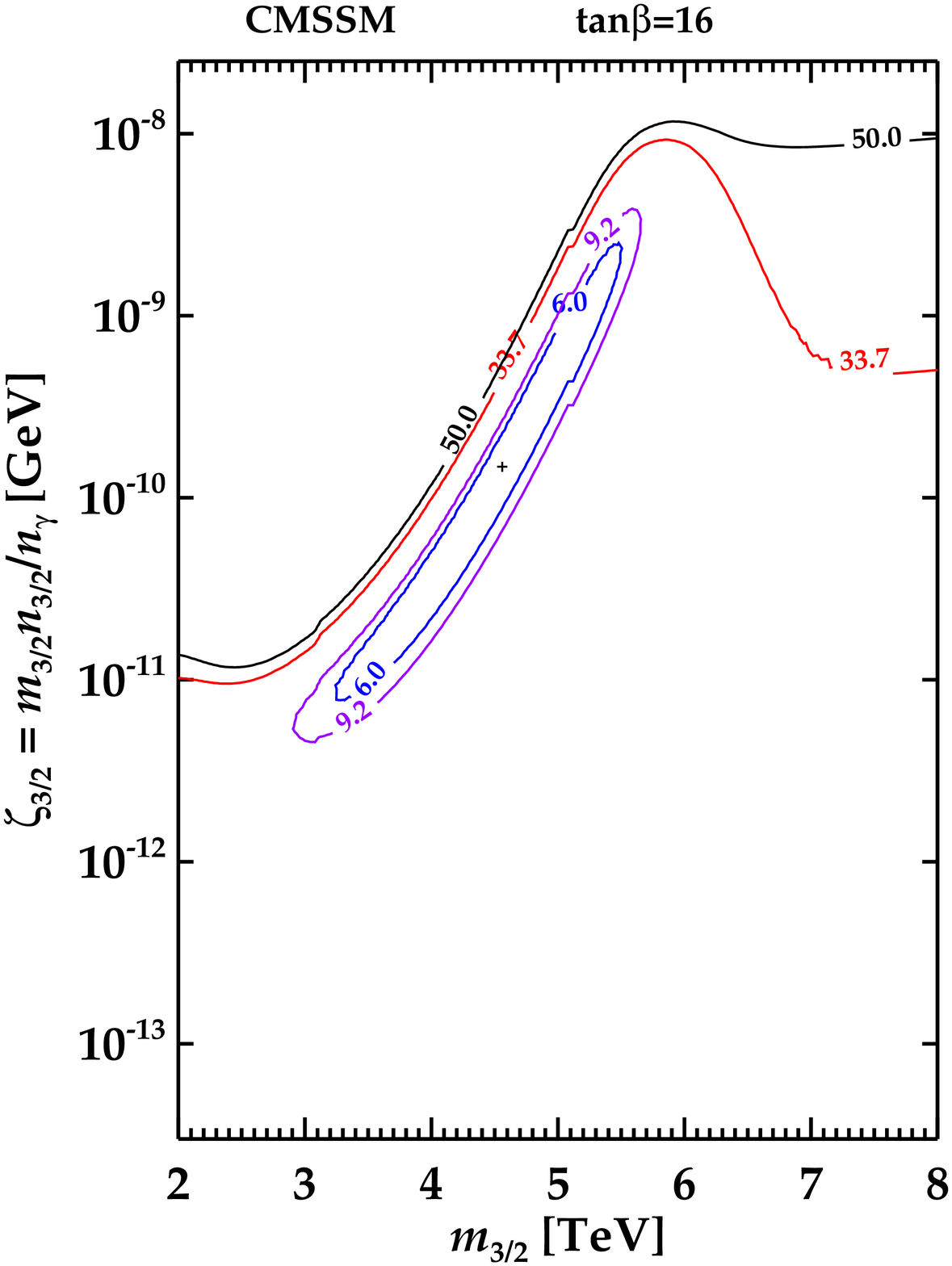,height=8cm}
\epsfig{file=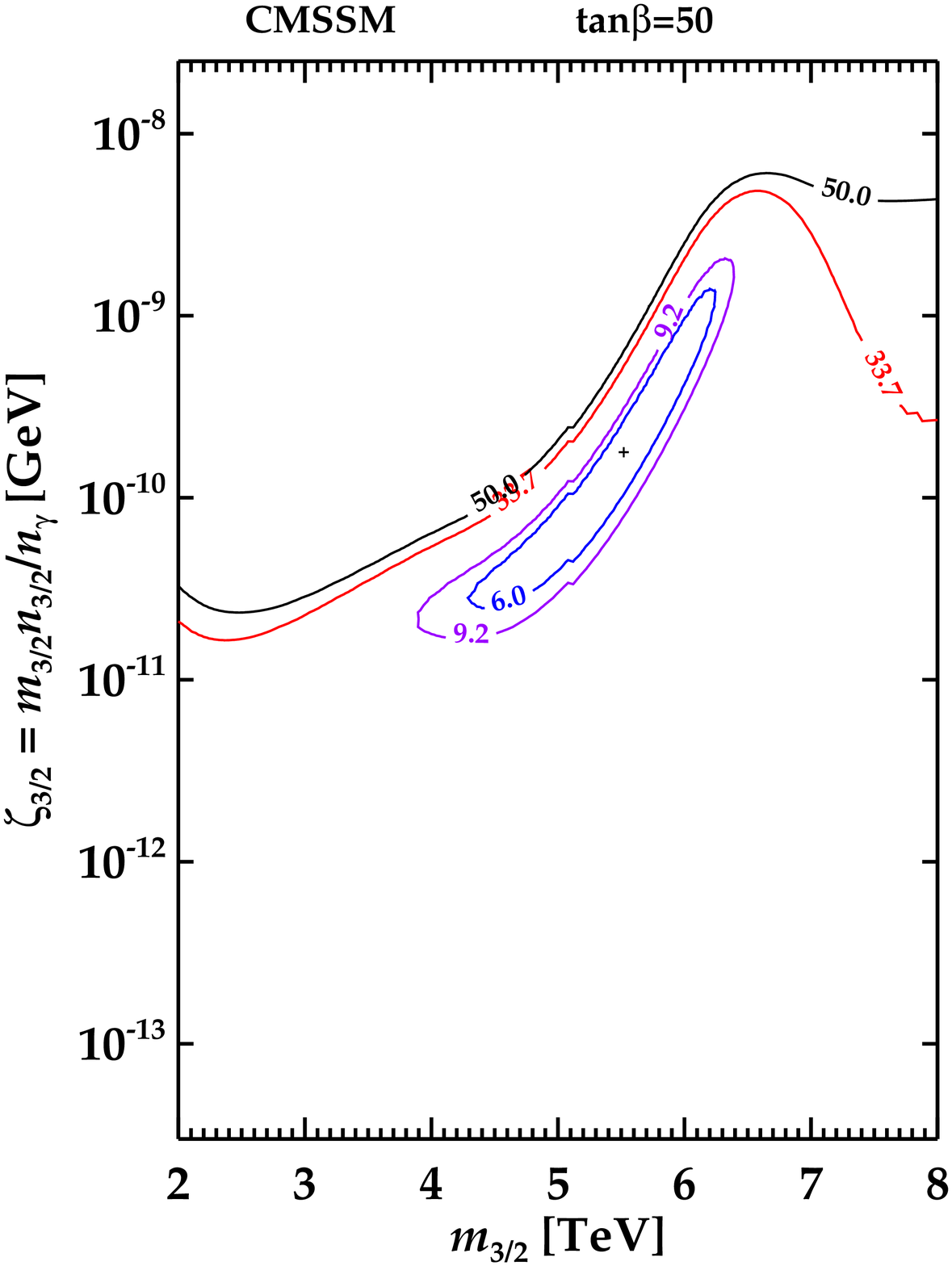,height=8cm}\\
\epsfig{file=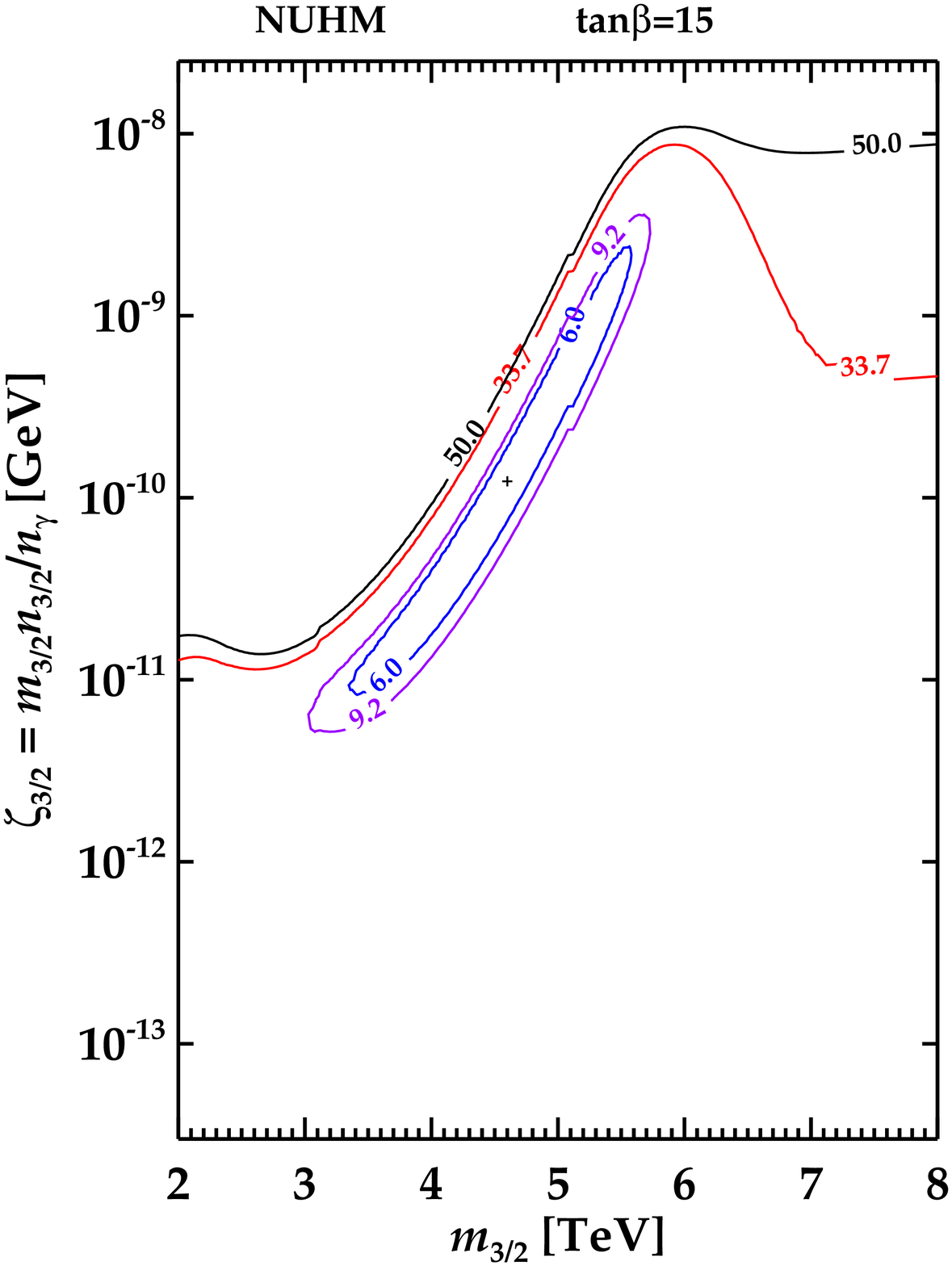,height=8cm}
\epsfig{file=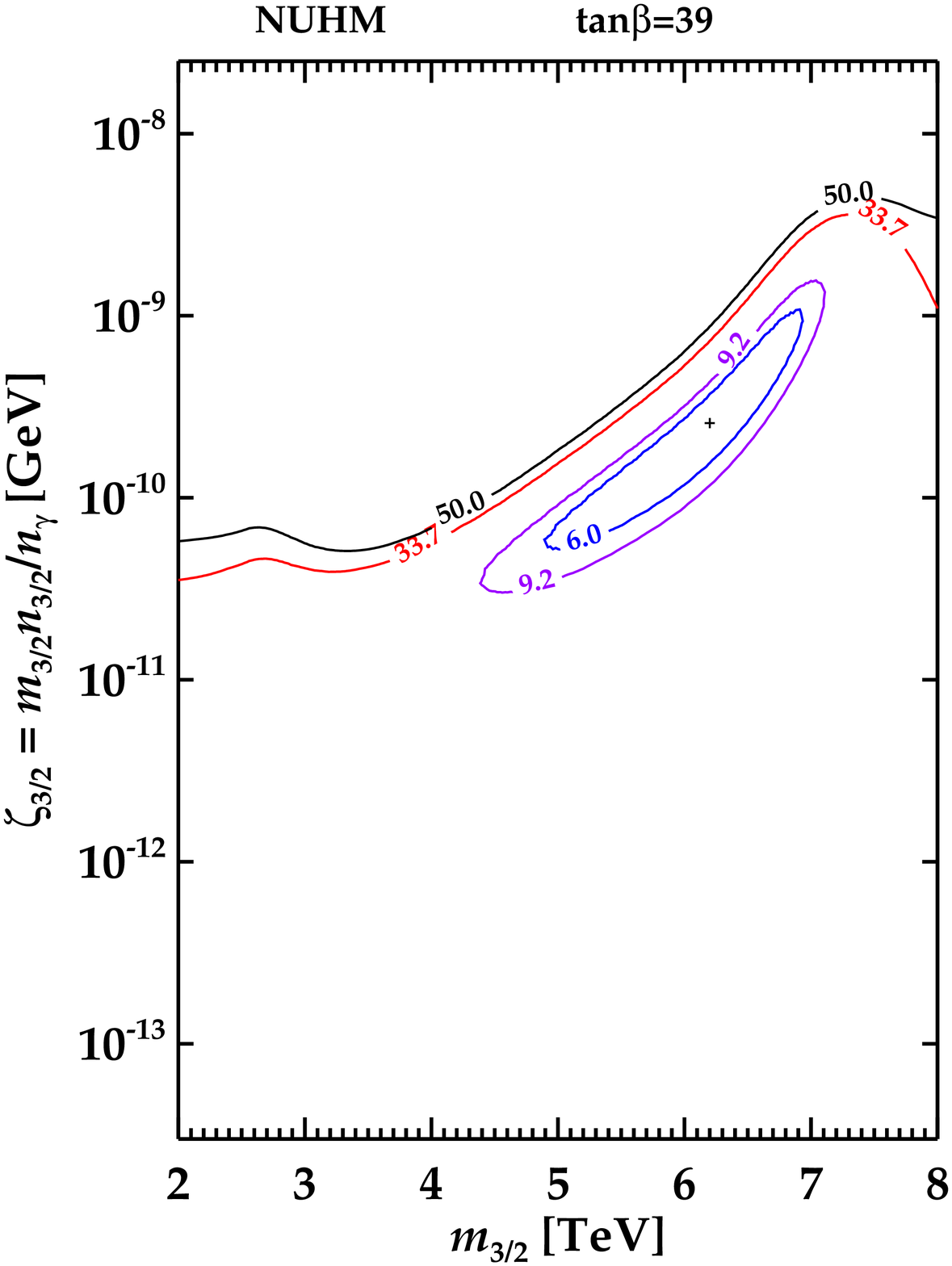,height=8cm}
\end{center}
\caption{\it
Contours of the $\chi^2$ function in the $(m_{3/2}, \zeta_{3/2})$ planes for the
models described in the first four rows of Table~\ref{tab:bestfits}:
$\chi^2 = 6.0$ (blue), $9.2$ (magenta), $33.7$ (red) and $50$ (black).
Also marked by black crosses are the best-fit points in the various models.}
\label{fig:chi2}
\end{figure}

As a refinement of this analysis of the compatibility between the
different light-element constraints, we display in Fig.~\ref{fig:chi2}
contours of the global $\chi^2$ function in the $(m_{3/2}, \zeta_{3/2})$ planes for the
models listed in the first four rows of Table~\ref{tab:bestfits}. 
The value of $\chi^2$ is determined from 
\beq
\chi^2 \; \equiv \; \left(\frac{Y_p - 0.2534}{0.0083} \right)^2 + \left(\frac{{\rm D/H} - 3.01 \times 10^{-5}}{0.27 \times 10^{-5}}\right)^2 +
 \left(\frac{{\li7/\rm H} - 1.23 \times 10^{-10}}{0.71 \times 10^{-10}}\right)^2
 + \left(\frac{\Omega_\chi^{(3/2)} h^2}{0.0045}\right)^2 ,
 \label{chi2}
 \eeq
 where 
 \beq
 \Omega_\chi^{(3/2)} = \frac{m_\chi}{m_{3/2}} \frac{n_\gamma}{\rho_c} \zeta_{3/2}\, 
\eeq
is the density of neutralinos produced in gravitino decays, $\rho_c$ being the critical density of the Universe~\footnote{We note
that it would in principle be possible to adapt the parameters of the models listed in Table~\ref{tab:bestfits} so that
the density of thermally-produced neutralinos is below the central WMAP value, in which case the constraint on
$ \Omega_\chi^{(3/2)}$ would be relaxed. However, since this constraint is not important in the neighbourhoods
of the best-fit points, we have not explored this option.}. 
In each case, we see a
narrow valley where $\chi^2 < 6.0$ (blue line), representing an acceptable
solution to the cosmological \li7 problem. In each case, the rise to $\chi^2 = 9.2$ 
(magenta line) is quite rapid, but the further
rise at smaller $m_{3/2}$ and larger $\zeta_{3/2}$ (due to a combination of
several constraints)
is much more rapid than that at 
larger $m_{3/2}$ and smaller $\zeta_{3/2}$ (where it is due solely to the cosmological \li7
problem). At high $\zeta_{3/2}$ the WMAP constraint plays a role in closing the contours
for $\chi^2 = 6$ and 9.2 as well as the bend and flattening of the $\chi^2 = 33.7$ and 50
contours respectively. The black crosses in Fig.~\ref{fig:chi2} indicate the best-fit points in the different
scenarios: their properties ($m_{3/2}$, $\zeta_{3/2}$, $\tau_{3/2}$, $\chi^2_{\rm min}$) are given in the
first four rows of Table~\ref{tab:bestfits}. We see that the best-fit gravitino mass varies 
between 4.6 and 6.2~TeV, and its abundance in the narrow range
between 1.0 and $2.6 \times 10^{-10}$~GeV.  The best-fit gravitino lifetimes fall in an even more
narrow range, $\tau_{3/2} \sim 210-280$~sec.  Thus the models
all show a close similarity in
the best-fit gravitino
abundance and lifetime values.  This behavior is typical
of decaying particle models we and other have studied, and
indicates that these
parameters exert the most sensitive control on the abundances of the light elements.

All four of the models have a $\chi^2_{\rm min}$ of about 3, which represents a significant reduction from the 
SBBN value of 33.7~\footnote{The differences between these values and those found in~\cite{CEFLOS1.5}
are due to a combination of factors, including the updated D/H abundance as well as the revised
value of $\Omega_b h^2$ as well as the new supersymmetric model parameters.}. 
However, there is also a price to be paid, in that we have 
added 2 parameters ($m_{3/2}$ and $\zeta_{3/2}$) to achieve that benefit.
As a result we have also decreased the number of degrees of freedom from 3 to 1.
Nevertheless, the fit probability for $\chi^2$ per degree of freedom of 3/1 is 
far superior to that for 33.7/3.

As also seen in Table~\ref{tab:bestfits}, we have explored the effects of
late-decaying massive gravitinos in six other scenarios: three other CMSSM
and NUHM1 scenarios and three subGUT scenarios, all of them consistent
with the current LHC and other constraints and representing different ways
in which the relic $\chi$ density can be brought into the range favoured by
WMAP and other experiments, as discussed earlier. 
The main features of our results for these scenarios are similar to
those of the previous models shown and we list 
the parameters of the best fits in the 5th to the 10th row of
Table~\ref{tab:bestfits}. Each of them lies within the ranges found in the
models discussed above.

\begin{figure}[h!]
\begin{center}
\epsfig{file=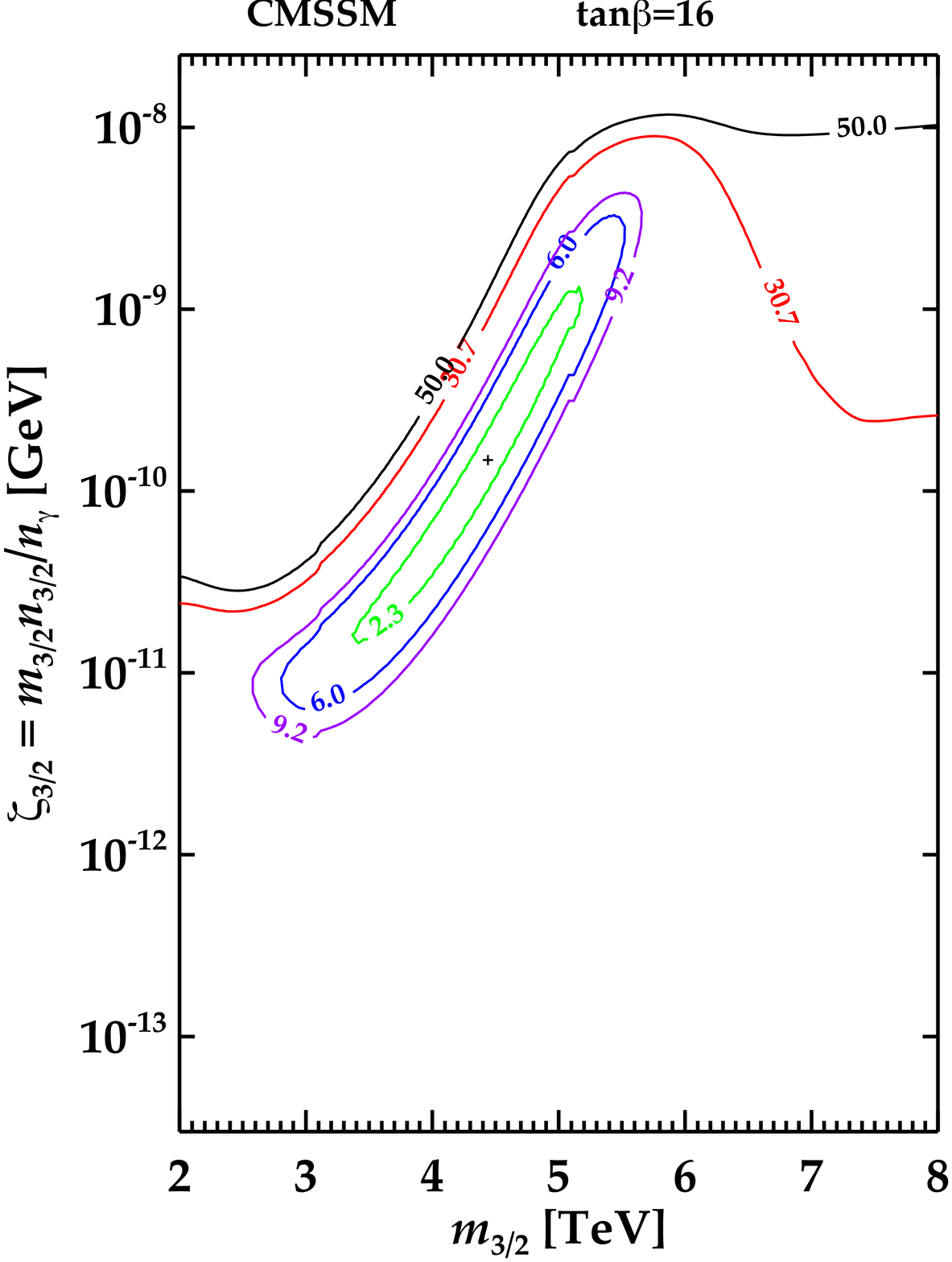,height=7cm}
\epsfig{file=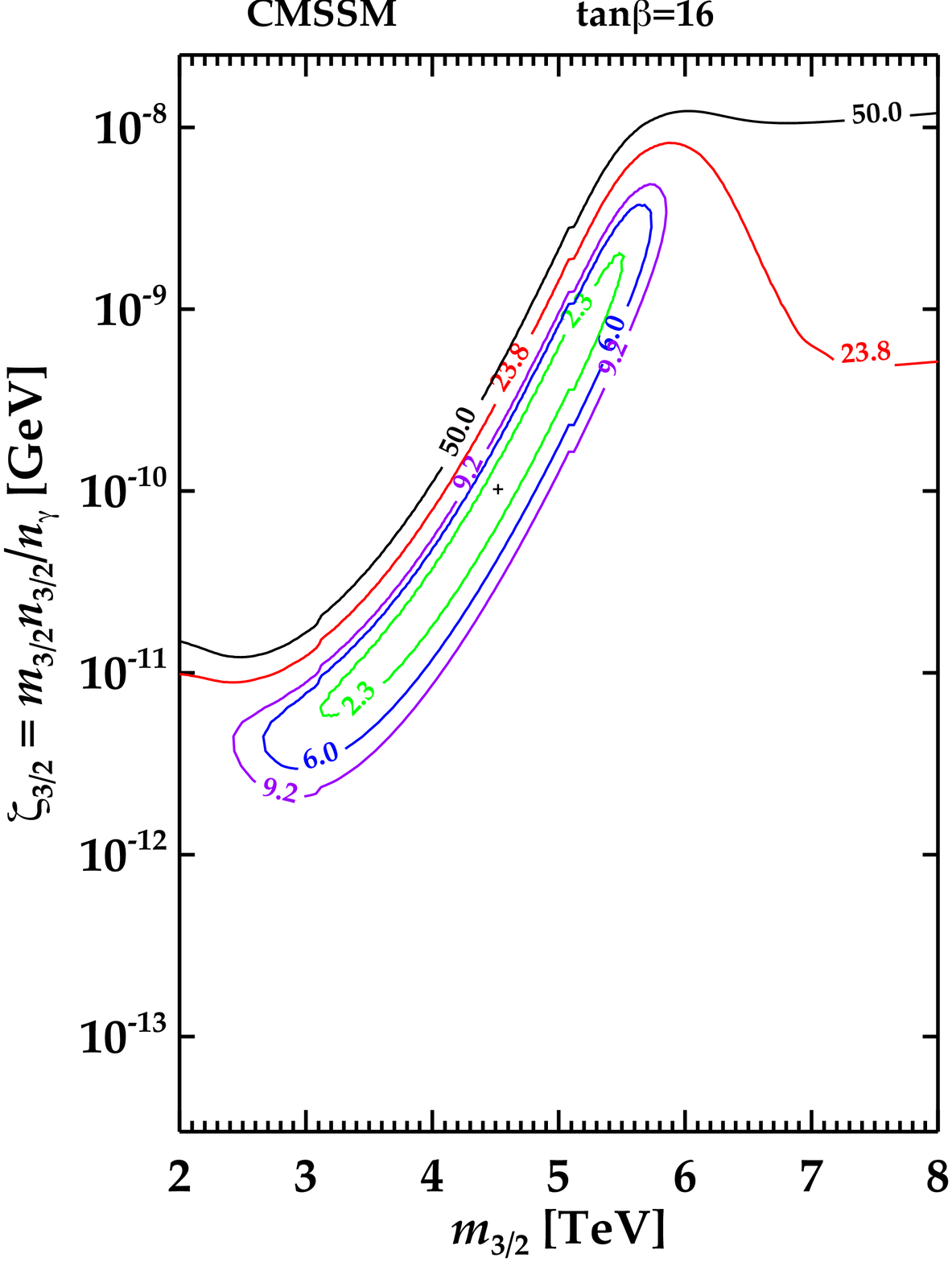,height=7cm}
\epsfig{file=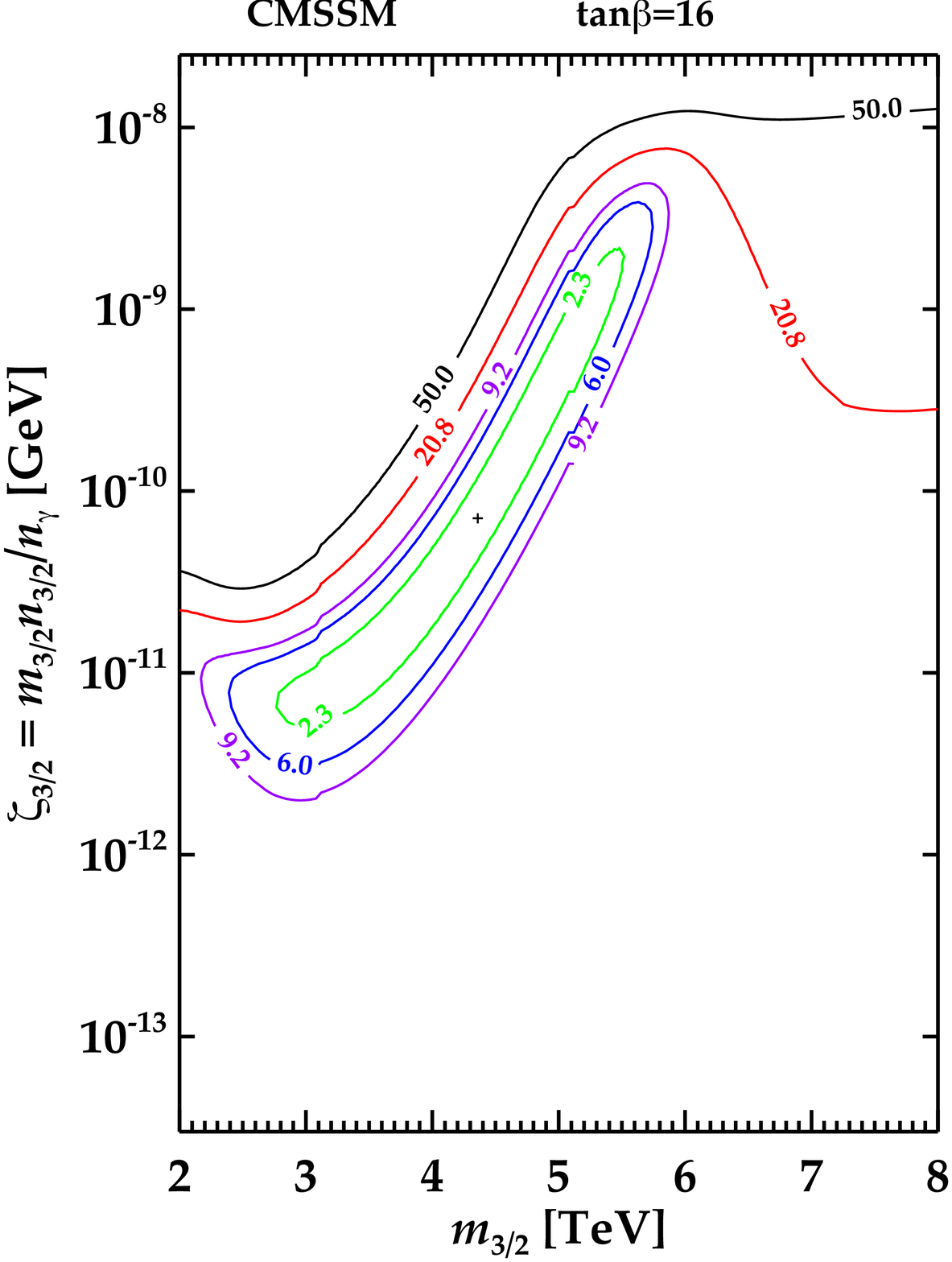,height=7cm}\\
\end{center}
\caption{\it
Contours of the $\chi^2$ function in the $(m_{3/2}, \zeta_{3/2})$ planes for the
low-mass CMSSM model described in the first row of Table~\ref{tab:bestfits} under
different assumptions on the D/H and \li7/H abundances. In the left panel we relax
the D/H constraint by using an increased uncertainty $\pm 0.70 \times 10^{-5}$, in
the middle panel we use the globular cluster value for \li7/H $= (2.34 \pm 0.05) \times 10^{-10}$,
and in the right panel we combine these assumptions. The contours are coloured as follows:
$\chi^2 = 2.3$ (green), $\chi^2 = 6.0$ (blue), $9.2$ (magenta), and $50$ (black).  The red contour corresponds to the SBBN $\chi^2$ for each case,
namely 30.7, 23.8, and 20.8 for the left, middle, and right panels 
respectively.
Also marked by black crosses are the best-fit points under these different assumptions.}
\label{fig:variouschi2}
\end{figure}

We have studied effects in the first of the models in Table~\ref{tab:bestfits}
under different assumptions on the D/H and \li7/H abundances, with the
results shown in Fig.~\ref{fig:variouschi2}. In the left panel we show
contours of the $\chi^2$ function if the D/H constraint (\ref{eq:D}) is
relaxed, keeping the same central value but assuming an uncertainty
of $\pm~0.70 \times 10^{-5}$, given by the sample variance. In this case, the banana-shaped
region with $\chi^2 < 6.0$ is significantly expanded, and the minimum
value of $\chi^2 = 1.25$ (compared with the SBBN value of 30.7). 
The middle panel of Fig.~\ref{fig:variouschi2}
shows results obtained keeping the more restrictive uncertainty in D/H
but assuming the globular cluster value for \li7/H $= (2.34 \pm 0.62) \times 10^{-10}$.
In this case, the `banana' is even broader, and the minimum value of $\chi^2 = 0.52$
(compared with the SBBN value of 23.8).
Finally, the right panel shows the effects of relaxing both the D/H and \li7/H
constraints, and we see that the combined result is more similar to the result
of relaxing the \li7/H constraint alone than that of relaxing the D/H constraint alone.
In this case the minimum value of $\chi^2 = 0.37$ (compared with 20.8 in SBBN), 
demonstrating again that
relaxing the \li7/H constraint is more important than relaxing the D/H constraint.

\begin{table*}[!tbh!]
\renewcommand{\arraystretch}{1.5}
\begin{center}
{
\begin{tabular}{|c|c||c|c|c|c||c|} \hline
ID &  $Y_p$ & $10^5 \ \rm D/H$ & $10^{10} \ \rm \li7/H$ & $\Omega_{\chi}^{(3/2)} h^2$ & $\chi^2_{\rm min}$ \\
\hline \hline
1 & 0.2487 & 3.27 & 2.12 & $5.0 \times 10^{-4}$ & 2.81 \\
2 & 0.2487 & 3.28 & 2.09 & $1.1 \times 10^{-3}$ &  2.86 \\
3 & 0.2487 & 3.26 & 2.14 & $4.4 \times 10^{-4}$ & 2.82 \\
4 & 0.2487 & 3.29 & 2.11 & $2.1 \times 10^{-3}$ & 3.14 \\
\hline
5 & 0.2487 & 3.32 & 2.01 & $6.5 \times 10^{-4}$ & 2.87 \\
6 & 0.2487 & 3.27 & 2.11 & $1.0 \times 10^{-3}$ & 2.86 \\
7 & 0.2487 & 3.29 & 2.08 & $4.7 \times 10^{-4}$ & 2.87 \\
8 & 0.2487 & 3.25 & 2.16 & $1.8 \times 10^{-3}$ & 2.96 \\
9 & 0.2487 & 3.31 & 2.04 & $1.2 \times 10^{-3}$ & 2.91 \\
10 & 0.2487 & 3.28 & 2.09 & $1.4 \times 10^{-3}$ & 2.89 \\
\hline
11 & 0.2487 & 3.55 & 1.63 & $5.1 \times 10^{-4}$ & 1.25 \\
12 & 0.2487 & 3.10 & 2.50 & $3.5 \times 10^{-4}$ & 0.52 \\
13 & 0.2487 & 3.15 & 2.40 & $2.5 \times 10^{-4}$ & 0.37 \\
\hline
\end{tabular}
 }
\caption{\it Predictions at the best-fit points for the models defined in Table~\ref{tab:bestfits}
for light-element abundances and the neutralino abundance $\Omega^{(3/2)}_\chi h^2$ arising from gravitino decays.
 }
\label{tab:bestabs}
\end{center}
\end{table*}

Table \ref{tab:bestabs} presents light-element abundances predicted at the best-fit points
for each of the models defined in Table~\ref{tab:bestfits}.
We see that all models select very similar best-fit values of D/H and \li7/H,
in line with the trends found in Fig.~\ref{fig:D_Li}, namely a combination of the 
highest tolerable D/H and lowest tolerable \li7/H.
We also see that, in all models, there is no change in the helium abundance
from its SBBN value.  
Finally, Table \ref{tab:bestabs} shows the best-fit prediction for the relic $\chi$ abundance due to
gravitino decays, $\Omega_{\chi}^{(3/2)} h^2$. These vary from $\sim (0.3-2) \times 10^{-3}$, and thus
are at most about half of the observational uncertainty in $\Omega_{\rm cdm} h^2$.  
The relic abundance constraint is therefore not a substantial contributor to the global $\chi^2$ function in the
regions of the best-fit points.

\section{Summary and Conclusions}

We had shown previously~\cite{CEFLOS1.5} that the late decays of massive gravitinos offered a possible
solution of the cosmological \li7 problem in a number of benchmark CMSSM scenarios
proposed before the start-up of the LHC. The subsequent non-discovery of supersymmetry
at the LHC~\cite{lhc}, the discovery of a Higgs boson~\cite{H}, the measurement of $B_s \to \mu^+ \mu^-$~\cite{bmm}
and the XENON100 dark-matter search experiment~\cite{XE100} have 
squeezed severely the allowed region of the CMSSM parameter space~\cite{MC8}, and
stimulating the exploration of alternative supersymmetric scenarios 
such as NUHM1 models with
non-universal Higgs masses and subGUT models in which soft supersymmetry-breaking
masses are assumed to be universal at some scale below the GUT scale~\cite{ELOS}.
In this paper, we have explored the most favoured sets of CMSSM and NUHM1 parameters not yet
excluded by the LHC~\cite{MC8}, as well as a number of alternative models within the NUHM1 and subGUT
frameworks, as summarized in Table~\ref{tab:bestfits}.

In contrast to the constraints on supersymmetric model parameters, the constraints on primordial
light-element abundances have remained essentially unchanged since~\cite{CEFLOS1.5}, as
summarized in Table~\ref{tab:abundances}. Thus the cosmological \li7 problem remains.

In this paper we have repeated the global $\chi^2$ likelihood analysis of~\cite{CEFLOS1.5},
and studied the regions of the gravitino parameters $(m_{3/2}, \zeta_{3/2})$ that offer the best
prospects of solving the cosmological \li7 problem. In all the models studied, we find global $\chi^2_{\rm min} \lappeq 3$
in the $(m_{3/2}, \zeta_{3/2})$ plane and therefore the cosmological \li7
problem can be regarded even more effectively solved than in~\cite{CEFLOS1.5}. 
Typical ranges of the gravitino parameters are
$4.6~{\rm TeV} < m_{3/2} < 6.2$~TeV and $1.0 \times 10^{-10}~{\rm GeV} < \zeta_{3/2} < 2.6 \times 10^{-10}~{\rm GeV}$.

In view of this persistence of the massive gravitino solution despite the impact of LHC results
on the supersymmetric parameter space, we conclude that the late decays of massive
gravitinos provide a robust solution to the cosmological \li7 problem.

\section*{Acknowledgments}

The work of R.H.C. was supported by the U.S. National Science Foundation Grant
PHY-08-22648 (JINA). The work of J.E. and F.L. was supported in
part by the London Centre for Terauniverse Studies (LCTS), using funding from the European
Research Council via the Advanced Investigator Grant 267352.
The work of B.D.F. was partially supported by the U.S. National Science Foundation Grant
PHY-1214082. The work of K.A.O. was supported in part by  
DOE grant DE-FG02-94ER-40823 at the University of Minnesota. The work of
V.C.S. was supported by Marie Curie International Reintegration grant SUSYDM-PHEN,
MIRG-CT-2007-203189.

\end{document}